\PassOptionsToPackage{table,xcdraw}{xcolor}
\documentclass[preprint,5p,times,twocolumn]{elsarticle}
\usepackage{amssymb}
\usepackage{color}
\usepackage{tabularx}
\usepackage{multirow}
\usepackage{amsmath}
\usepackage{booktabs}
\usepackage{subfigure}
\usepackage{threeparttable}
\usepackage{times}
\usepackage{xcolor}
\usepackage{longtable}
\usepackage{graphicx}      
\usepackage{makecell}    
\usepackage{pifont}
\usepackage{lscape}
\usepackage{ltablex}
\usepackage{multicol}
\usepackage[table,xcdraw]{xcolor}
\usepackage{textcomp}
\usepackage{xltabular}
\usepackage{nicematrix}
\usepackage{chemformula}

\usepackage{hyperref}
\hypersetup{hidelinks,
	colorlinks=true,
	allcolors=black,
	pdfstartview=Fit,
	breaklinks=true}

\newcolumntype{Z}{>{\centering\arraybackslash}X}
\makeatletter

\newcommand{\Rmnum}[1]{\expandafter\@slowromancap\romannumeral #1@}
\makeatother

\def\eg{\textit{e.g.}}

\journal{XXX}

\begin{document}

\begin{frontmatter}



\title{Taming Vision-Language Models for Medical Image Analysis: A Comprehensive Review} 


\author[mymainaddress]{Haoneng Lin} 

\author[mymainaddress]{Cheng Xu\corref{mycorrespondingauthor}}
\cortext[mycorrespondingauthor]{Corresponding authors}
\ead{cs-cheng.xu@polyu.edu.hk.}


\author[mymainaddress]{Jing Qin}

\address[mymainaddress]{Center of Smart Health, Hong Kong Polytechnic University, Hong Kong, China}


\begin{abstract}
Modern Vision-Language Models (VLMs) exhibit unprecedented capabilities in cross-modal semantic understanding between visual and textual modalities. 
Given the intrinsic need for multi-modal integration in clinical applications, VLMs have emerged as a promising solution for a wide range of medical image analysis tasks. 
However, adapting general-purpose VLMs to medical domain poses numerous challenges, such as large domain gaps, complicated pathological variations, and diversity and uniqueness of different tasks.
The central purpose of this review is to systematically summarize recent advances in adapting VLMs for medical image analysis, analyzing current challenges, and recommending promising yet urgent directions for further investigations.     
%
%
We begin by introducing core learning strategies for medical VLMs, including pretraining, fine-tuning, and prompt learning. 
We then categorize five major VLM adaptation strategies for medical image analysis. 
These strategies are further analyzed across eleven medical imaging tasks to illustrate their current practical implementations. 
Furthermore, we analyze key challenges that impede the effective adaptation of VLMs to clinical applications and discuss potential directions for future research. 
%
%
We also provide an open-access repository of related literature to facilitate further research, available at \href{github}{https://github.com/haonenglin/Awesome-VLM-for-MIA}.
It is anticipated that this article can help researchers who are interested in harnessing VLMs in medical image analysis tasks have a better understanding on their capabilities and limitations, as well as current technical barriers, to promote their innovative, robust, and safe application in clinical practice.    
\end{abstract}





\end{frontmatter}




\section{Introduction}

%
Medical image analysis plays a crucial role in modern healthcare systems, supporting accurate and timely diagnosis, treatment planning, and prognosis~\cite{duncan2000medical, litjens2017survey}. 
In the past decade, deep learning has emerged as a powerful tool in medical image analysis, leading to revolutionary progress in a wide range of applications~\cite{shen2017deep, esteva2019guide}. 
However, most traditional approaches focused solely on images, neglecting that, in clinical practice, most procedures do not only rely on images, but also, probably more closely, depend on linguistic data, such as clinical notes, image reports, patients' historical records, and so on.   
To the end, while these models demonstrated compelling performance in well-defined tasks, like tumor detection~\cite{viswanathan2024towards} or organ segmentation~\cite{shi2022deep}, most of them cannot be effectively used or seamlessly integrated into clinical routines due to their inherent shortcomings in capturing multifaceted clinical contexts, which are indispensable for holistic disease understanding and precise decision-making~\cite{haug2023artificial}.      

%
%
This inherent multi-modal nature of clinical procedures has prompted increasing interest in combining vision and language data to enhance accuracy and efficacy. 
Conventional multi-modal fusion methods typically rely on hand-crafted architectures tailored to narrow tasks~\cite{perez2019mfas}, suffer from limited annotated datasets, and often face scaling bottlenecks.
Recent breakthroughs in Vision-Language Models (VLMs), such as CLIP~\cite{radford2021learning}, BLIP~\cite{li2022blip}, GPT-4V~\cite{yang2023dawn}, and Gemini~\cite{team2023gemini}, have introduced a transformative paradigm for vision-language cross-modal learning.
These models are pretrained on large-scale image-text pairs using contrastive and generative learning objectives, which facilitates them to acquire rich, modality-agnostic semantic embeddings. 
This unified representation of visual and textual information has empowered a range of applications, including zero-shot learning~\cite{lei2024ez, saha2024improved}, cross-modal retrieval~\cite{tankala2024wikido, yang2024rebalanced}, visual question answering~\cite{zeng2023x, sima2024drivelm}, and so on. 
%
%

Capitalizing on the unified visual-textual representations learned by VLMs, recent efforts have focused on their strategic integration into clinical workflows. 
However, adapting general-domain VLMs to the medical domain presents distinct challenges. 
First, the domain gap, manifested through differences in visual features and terminologies, hinders direct transfer. Although several medical VLMs (\eg, PubMedCLIP~\cite{eslami2023pubmedclip} and UniMed-CLIP~\cite{khattak2024unimed}) have been specifically developed to mitigate this gap, they often face limitations in domain-specific semantic alignment and insufficient integration of structured clinical knowledge.
Second, clinical applications demand extremely high reliability, yet current VLMs often exhibit vulnerability to subtle pathological variations. 
Third, their ``black-box'' mechanisms raise concerns about interpretability, transparency, and regulatory compliance in high-stakes settings. 


These challenges underscore the need for a systematic review of methodologies for adapting VLMs to medical domains.
Although several recent reviews~\cite{liu2024visual,li2025vision,lu2025integrating} have outlined VLM advancements in medical domain, they primarily emphasize model architectures, benchmark comparisons, or deployment pipelines, without systematically analyzing the methodological pathways for adapting general-purpose VLMs to diverse, task-specific medical applications. 
In contrast, our review presents a unified framework of five adaptation strategies that captures the breadth of current research while offering a structured lens for understanding and advancing medical VLM customization. 
Specifically, we present a comprehensive analysis of medical VLM adaptation through five key dimensions: (1) a comprehensive review of existing VLMs training paradigms applied to medical image analysis; (2) a structured taxonomy of task-specific adaptation strategies tailored for medical applications, encompassing input augmentation~\cite{ni2024m2trans, abacha20233d}, feature enrichment~\cite{zhong2023ariadne,li2024tp}, supervision enhancement~\cite{lanfredi2025enhancing, chen2023bomd}, task-specific fine-tuning~\cite{chen2024can, yin2024histosyn}, and direct deployment~\cite{ghosh2024clipsyntel,ferber2024context}; (3) a detailed analysis of how these adaptation strategies are instantiated across eleven representative medical image analysis tasks, including medical report generation, visual question answering (VQA), image segmentation, anomaly detection, image classification, image synthesis, image-text retrieval, super-resolution, image registration, object detection, and multi-task learning; (4) an in-depth discussion of the core challenges associated with adapting VLMs to clinical scenarios, such as data scarcity, modality misalignment, and computational constraints; and (5) a forward-looking outlook on emerging research directions and practical design considerations to guide future developments in this field.
Through this structured lens, we aim to provide actionable insights for developing scalable, robust, and clinically trustworthy VLM systems for medical image analysis. 

\begin{figure*}[!t]
\centering
\includegraphics[width=\textwidth]{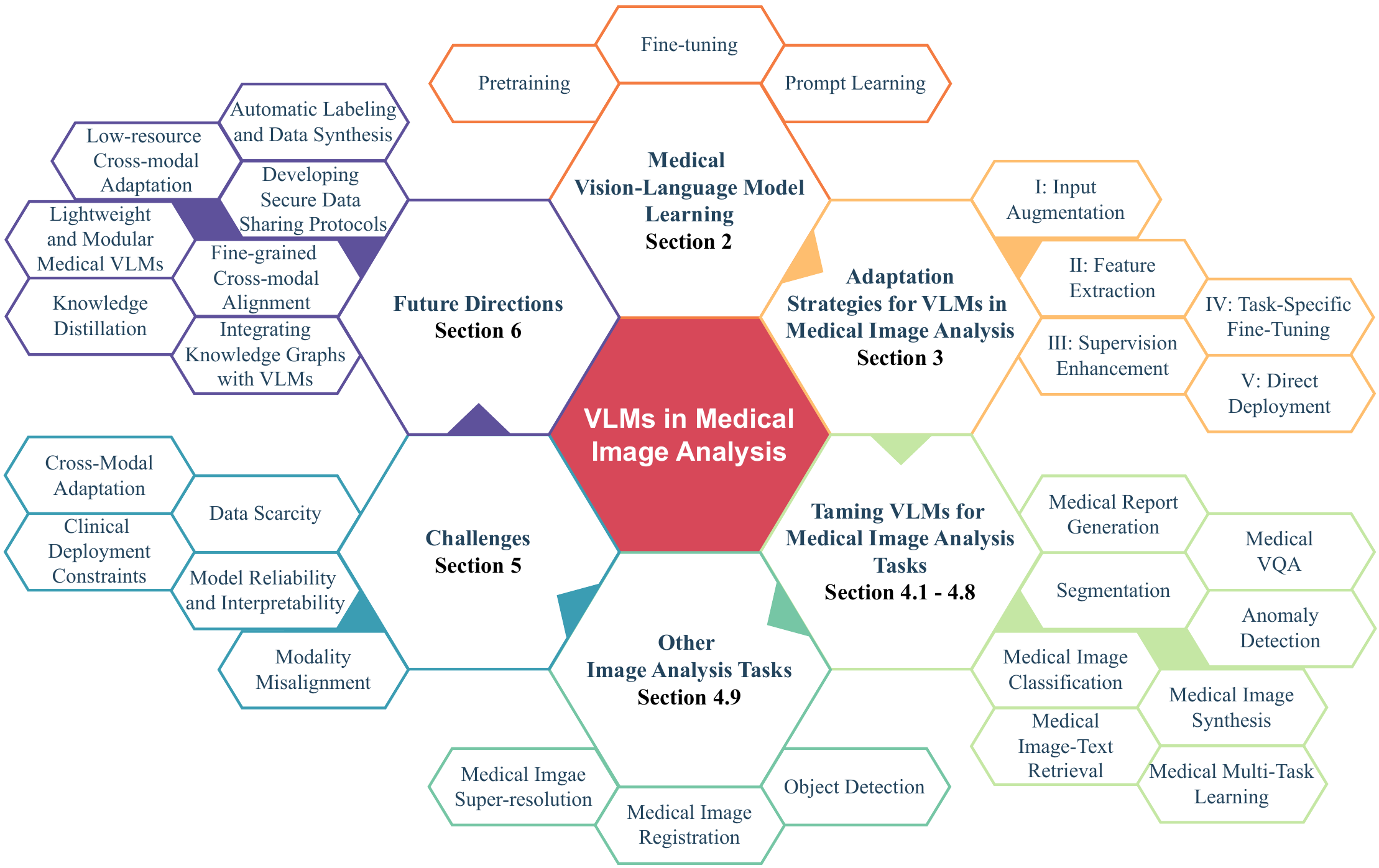}
\caption{The structure and main contents of this review.}
\label{Fig1}
\end{figure*}

Fig.~\ref{Fig1} illustrates the overall structure and main contents of this survey. The remainder of this paper is organized as follows. 
Section 2 introduces the learning paradigms of existing VLMs applied to the medical domain, including pretraining techniques, fine-tuning strategies, and prompt learning approaches. Section 3 presents a systematic taxonomy of adaptation strategies for downstream medical tasks, categorized into five principal approaches: input augmentation, feature extraction, supervision enhancement, task-specific fine-tuning, and direct deployment. Section 4 reviews how each adaptation strategy is instantiated across a wide range of medical tasks. Methodological patterns and design choices are analyzed within each task category. Section 5 discusses major challenges associated with VLM adaptation in medical image analysis. Section 6 outlines promising future research directions.
Finally, the paper concludes with a summary of key insights and identifies potential pathways for advancing the development and clinical integration of VLMs in medical image analysis.

\section{Medical Vision-Language Model Training}
\label{sec:VLM learning}
The advent of VLMs has catalyzed significant changes in multi-modal learning paradigms for medical image analysis. 
These models have been used in a wide range of tasks, such as classification, segmentation, report generation, VQA, etc.
Table~\ref{tab:VLM_uasge} summarizes existing VLMs for medical image analysis. 
To adapt VLMs to medical domains, nowadays, researchers mainly focus on three training schemes, pretraining, fine-tuning, and prompt learning. 
%
This section will review these training schemes, discussing their merits and shortcomings. 

\subsection{Pretraining}

Recent advances in VLMs have markedly improved performance on various natural image tasks. These models leverage large-scale pretraining techniques to learn multi-modal representations from image-text pairs, facilitating downstream applications such as zero-shot classification, image-text retrieval, and VQA. Despite their success in natural domains, applying vision-language pretraining (VLP) to medical imaging presents unique challenges, including data scarcity and complex cross-modal alignment.


\textbf{Data Scarcity}. UniMed-CLIP~\cite{khattak2024unimed} constructs UniMed, a large-scale multi-modal medical image-text dataset comprising X-ray, CT, MRI, ultrasound, pathology, and fundus images. Using large language models to generate image-text pairs from classification datasets, UniMed-CLIP~\cite{khattak2024unimed} enables scalable multi-modal VLP, significantly enhancing zero-shot generalization across different imaging modalities.

VQA-driven VLM pretraining has emerged as a promising strategy for guiding models toward clinically relevant features. For example, VQA-Pretraining~\cite{su2024design} constructs multi-granular question-answer pairs from medical reports and incorporates a quasi-textual feature transformer to bridge the vision-language gap. This approach significantly enhances the performance of report generation, classification, segmentation, and detection tasks by improving modality alignment.

\textbf{Cross-model Alignment}. Effective vision-language pretraining hinges on the ability to align medical images with textual descriptions. However, many existing VLM pretraining approaches fail to capture the fine-grained semantics in radiology reports, leading to suboptimal cross-modal representations. To address this, MPMA~\cite{zhang2023multi} proposed a multi-task paired masking with alignment (MPMA) framework that integrates a global and local alignment (GLA) module within a joint image-text reconstruction paradigm. This strategy enhances cross-modal interactions, leading to better semantic consistency between medical images and their textual descriptions during pretraining.

MedKLIP~\cite{wu2023medklip} incorporates a triplet extraction module to structure textual inputs, effectively reducing syntactic complexity and enhancing supervision signals in VLM pretraining. By leveraging knowledge base querying (entity translation), MedKLIP strengthens the relationships between medical entities in the vision-language embedding space. Similarly, Zhang et al.~\cite{zhang2023knowledge} introduces a knowledge-boosted contrastive learning framework, integrating unbiased open-set knowledge representations to mitigate negative sample noise in contrastive VLP. This approach improves the semantic consistency between vision-language mutual information and clinical knowledge, demonstrating superior zero-shot and few-shot performance across classification, segmentation, and retrieval tasks.

PTUnifier~\cite{chen2023towards} introduces a soft prompt-based unification strategy for VLM pretraining, allowing a single model to process image-only, text-only, and image-text inputs seamlessly. This approach improves the model's versatility, achieving state-of-the-art (SOTA) performance on image classification, text summarization, and VQA.

M-FLAG~\cite{liu2023m} proposes a novel frozen language model-based pretraining technique, incorporating an orthogonality loss to optimize latent space geometry. This method reduces model parameters by 78\% while significantly enhancing performance across classification, segmentation, and object detection tasks. Meanwhile, UniChest~\cite{dai2024unichest} introduces a conquer-and-divide pretraining strategy, which improves multi-source dataset integration by balancing shared feature learning and personalized query networks, ensuring robust task performance across multiple CXR datasets.

Contrastive learning has been widely adopted in medical VLP to enhance discriminative power in image-text alignment. For instance, CXR-CLIP~\cite{you2023cxr} integrates two contrastive losses to optimize study-level representation learning for chest X-ray image-text pairs, leading to improved disease classification. Similarly, MaCo~\cite{huang2024enhancing} introduces masked contrastive learning to refine fine-grained image representation learning. By implementing a correlation-weighted mechanism, MaCo adjusts masked X-ray image patches and their corresponding reports during VLM pretraining, enabling strong zero-shot generalization across multiple medical tasks.

Clinical-BERT~\cite{yan2022clinical} introduces a multi-task pretraining framework that explicitly incorporates medical knowledge by leveraging domain-specific objectives such as Clinical Diagnosis (CD), Masked MeSH Modeling (MMM), and Image-MeSH Matching (IMM). These tasks enhance the model’s ability to associate radiographic images with medical terminology, leading to improved radiograph diagnosis and report generation across multiple datasets. Meanwhile, RET-CLIP~\cite{du2024ret} employs a CLIP-style contrastive learning strategy to pretrain a retinal image foundation model on a dataset of 193,865 patients, optimizing representations at the left eye, right eye, and patient levels. This large-scale pretraining enables RET-CLIP to achieve state-of-the-art performance across eight datasets, demonstrating its strong generalization ability in diabetic retinopathy detection, glaucoma classification, and multi-disease diagnosis.

Additionally, medical images, particularly high-resolution 3D scans (e.g., CT, MRI), pose significant computational challenges for VLM pretraining due to hardware constraints and potential information loss from downsampling. Traditional 2D VLM methods struggle to handle volumetric data effectively. To address this, T3D~\cite{liu2023t3d} introduced the first VLP framework tailored for high-resolution 3D medical images, integrating text-informed contrastive learning and image restoration tasks. This approach allows the model to learn 3D representations without losing critical anatomical details, significantly improving tumor segmentation and disease classification.

\subsection{Fine-tuning}

Fine-tuning pretrained VLMs is another important way for adapting them to medical applications, as direct deployment of pretrained VLMs often results in suboptimal performance due to domain shifts. However, full fine-tuning of large medical VLMs is computationally expensive and may negatively impact generalization. To address these challenges, 
current fine-tuning methods for medical VLMs can be broadly categorized into three key areas:
1) Efficient fine-tuning methods, such as Parameter-Efficient Fine-Tuning (PEFT)~\cite{fu2023effectiveness}, to improve training efficiency while maintaining performance.
2) Domain-specific text-based fine-tuning, ensuring that fine-tuned models align with domain-specific terminology and structured pathology reports.
3) Robustness-oriented fine-tuning, improving model stability in noisy and adversarial environments. These studies reflect a broader trend towards more efficient, adaptive, and robust fine-tuning techniques, enabling scalable deployment of medical VLMs across diverse clinical applications.

\textbf{Parameter-Efficient Fine-Tuning (PEFT) for Medical VLMs}. With the increasing scale of VLMs, traditional full fine-tuning methods have become computationally prohibitive. Instead, Parameter-Efficient Fine-Tuning (PEFT) techniques have been developed to adapt large models with minimal computational overhead while preserving their generalization capabilities. Chen et al.~\cite{chen2024can} investigate whether fine-tuning techniques designed for LLMs can be effectively transferred to medical VLMs. This study evaluates the impact of various fine-tuning strategies on training efficiency and downstream performance in the medical domain. Their findings suggest that certain PEFT techniques, such as low-rank adaptation (LoRA) and adapter-based tuning, can significantly reduce computational costs while maintaining strong performance in medical image-text tasks.

\textbf{Domain-specific Fine-tuning}. Beyond computational efficiency, domain-specific adaptation is critical for fine-tuning medical VLMs, especially when dealing with structured clinical reports and pathology documentation. Since medical terminology differs significantly from general texts, VLMs require fine-tuning on domain-specific corpora to improve their semantic understanding and diagnostic accuracy. Kraišniković et al.~\cite{kraivsnikovic2025fine} fine-tune a BERT-based model for German-language pathology reports, demonstrating that domain-specific fine-tuning improves contextual representation learning in diagnostic text analysis. Their approach integrates explainable artificial intelligence (XAI) to enhance interpretability, making the fine-tuned model more reliable for clinical decision support. This study underscores the importance of language-specific fine-tuning for medical applications, as many pretrained VLMs are predominantly trained on English datasets, limiting their effectiveness in multilingual healthcare scenarios. 

\textbf{Robustness-oriented Fine-tuning}. Deploying fine-tuned medical VLMs in real-world clinical applications requires ensuring their robustness to adversarial noise. Medical images and clinical texts often contain inconsistent or noisy data, which can degrade model performance and introduce uncertainty in medical decision-making. To tackle this issue, Han et al.~\cite{han2024light} proposed the Rectify Adversarial Noise (RAN) framework, which incorporates multi-modal adversarial defense mechanisms into medical VLM fine-tuning. Their study finds that moderate levels of adversarial noise can enhance model robustness, whereas excessive noise leads to performance degradation. This research highlights a crucial consideration in medical VLM fine-tuning: while enhancing generalization, fine-tuned models must also be evaluated under real-world noisy conditions to ensure stability. Integrating adversarial training techniques during fine-tuning can improve model reliability, making them more trustworthy for clinical deployment.

\subsection{Prompt Learning}


Although fine-tuning significantly reduces computational overhead compared to pretraining, it still remains computationally expensive and often requires complex task-specific modifications to adapt large-scale VLMs for medical applications. To address this issue, prompt learning has emerged as an efficient alternative, allowing models to leverage pretrained knowledge with minimal modifications. Prompt learning for medical VLMs (Med-VLMs) has been explored in three primary directions: 1) Task-driven prompt optimization for adapting foundation models to medical imaging. 2) Enhancing robustness and generalization across different datasets and noisy environments. 3)Improving explainability and integrating external medical knowledge for clinically reliable predictions.

\textbf{Task-Driven Prompt Optimization}.
Recent studies have explored prompt learning techniques to enhance the robustness, interpretability, generalizability, and adaptability of medical vision-language models (Med-VLMs) in low-resource settings.

One of the earliest challenges in adapting foundation models to medical imaging is their limited out-of-the-box performance compared to task-specific deep learning methods. For example, while models like Segment Anything Model (SAM) exhibit strong generalization in natural image segmentation, their performance in medical image segmentation is often underwhelming due to a lack of domain-specific adaptation. To address this, Task-Driven Prompt Evolution (SAMPOT)~\cite{sathish2023task} introduces a prompt optimization framework that refines human-provided prompts based on downstream segmentation performance. Unlike static prompt engineering, SAMPOT dynamically adjusts prompts to improve segmentation accuracy in lung X-ray images, leading to a 75\% improvement over initial prompts. Beyond segmentation, prompt learning has been leveraged for multi-modal image recognition. Multi-modal Recursive Prompt Learning (MmRPL)~\cite{jia2024multi} improves hierarchical vision-text fusion by introducing recursive prompt updates across network layers. The integration of mixup embedding techniques further enhances domain generalization, making MmRPL highly effective in base-to-novel classification and domain-adaptive zero-shot learning. These studies demonstrate the effectiveness of adaptive prompt learning over traditional handcrafted prompts.

\textbf{Enhancing Robustness and Generalization of Med-VLMs}.
Prompt learning has also been employed to improve the robustness of Med-VLMs against noise and their generalization across datasets. PromptSmooth~\cite{hussein2024promptsmooth} introduces an adaptive textual prompt learning framework that enables Med-VLMs to maintain high accuracy under varying levels of Gaussian noise while significantly reducing computational costs. Similarly, Domain-Controlled Prompt Learning~\cite{cao2024domain} proposes a large-scale specific domain foundation model (LSDM) with domain bias control, allowing Med-VLMs to achieve superior performance in cross-dataset generalization. Furthermore, a comprehensive study on medical image understanding~\cite{qin2022medical} reveals that well-designed medical prompts can effectively elicit knowledge from pretrained VLMs, improving performance in zero-shot and few-shot learning scenarios.

\begin{table*}[!t]\footnotesize
\centering
\caption{Summary of existing VLMs utilized/developed for various medical image analysis tasks.
For readers' efficient reference, we listed their release years, datasets and scales (M: million, B: billion), training domains (N: nature, M : medical), and applicable tasks.}
\label{tab:VLM_uasge}
\setlength{\tabcolsep}{5pt}
\scalebox{0.86}{
\begin{tabular}{@{}lccccccccccccc@{}} 
\toprule
\multicolumn{5}{c}{\textbf{Model Information}} & \multicolumn{9}{c}{\textbf{Tasks}} \\
\cmidrule(r){1-5} \cmidrule(l){6-14}
\textbf{Models} & \textbf{Date} & \textbf{Datasets} & \textbf{Dataset Scales} & \textbf{N\textbackslash M} & \textbf{MRG} & \textbf{VQA} & \textbf{SEG} & \textbf{AD} & \textbf{MIC} & \textbf{MIS} & \textbf{MTIR} & \textbf{MMTL} & \textbf{Other}\\


\hline

VL-BERT~\cite{su2019vl} & 2020 & \makecell[c]{BooksCorpus\\English Wikipedia\\Conceptual Captions} & 3.3M image-text pairs & N & \ding{51} &  &  &  &  &  &  &  &  \\ 

\rowcolor{gray!30}CLIP~\cite{radford2021learning} &2021&WebImageText&400M image-text pairs&N&&\ding{51}&\ding{51}&\ding{51}&\ding{51}&\ding{51}&\ding{51}&\ding{51}&\ding{51}	\\	

GLIDE~\cite{nichol2021glide}&2021 &Self-Constructed &250 M image-text pairs &N&&&&&&\ding{51}&&& \\		

\rowcolor{gray!30}BLIP~\cite{li2022blip} &2022 & \makecell[c]{Conceptual Captions\\LAION,Visual Genome\\COCO, SBU, Conceptual 12M~} & 129M image-text pairs&N&&\ding{51}&\ding{51}&&&&&&\ding{51} \\	

GLIP~\cite{li2022grounded}&2022&\makecell[c]{SBU,COCO\\O365, FourODs, Cap4M\\CC3M, GoldG, Cap24M}&27M  image-text pairs&N&&&\ding{51}&&&&&&\ding{51}	\\				
\rowcolor{gray!30}Stable Diffusion~\cite{rombach2022high}&2022&\cellcolor{gray!30}N/A&\cellcolor{gray!30}N/A&N&&&&&\ding{51}&\ding{51}&&&\\				

ALBEF~\cite{li2021align}&2023&\makecell[c]{Visual Genome\\COCO, Conceptual  12M\\SBU, Conceptual Captions}&14.1M image-text pairs&N&\ding{51}&&&&&&&&	\\	

\rowcolor{gray!30}BLIP-2~\cite{li2023blip}&2023&\cellcolor{gray!30}\makecell[c]{~CC3M, COCO, LAION400M~~\\SBU, CC12M, Visual Genome}&\cellcolor{gray!30}129M image-text pairs&N&&\ding{51}&&&&&&&	\\	

Caption Anything~\cite{wang2023caption}&2023&N/A&N/A&N&&&&&&&&&\ding{51}	\\

\rowcolor{gray!30}DALL·E 3~\cite{betker2023improving}&2023&N/A&N/A&N&&&&&&&&&\ding{51}	\\		

Dreamlike Photoreal  &2023&N/A&N/A&N&&&&&&&&&\ding{51}		\\			

\rowcolor{gray!30}EVA-CLIP~\cite{sun2023eva}&2023&Merged-2B&2B image-text pairs&N&&\ding{51}&&&&&&\ding{51}&		\\

Gemini 1.0~\cite{team2023gemini}&2023&N/A&N/A&N&&\ding{51}&&&\ding{51}&&&&		\\	

\rowcolor{gray!30}Grounding DINO-T~\cite{liu2024grounding}&2023&\cellcolor{gray!30}\makecell[c]{RefC, OI, O365\\~~~~~~~COCO, GoldG, CC3M~~~~~~~}&\cellcolor{gray!30}N/A&N&&&&\ding{51}&&&&&		\\

MiniGPT-4~\cite{zhu2023minigpt}&2023&\makecell[c]{SBU, LAION\\Conceptual Captions}&5M image-text pairs&N&\ding{51}&&&&&&&&		\\

\rowcolor{gray!30}MiniGPT-v2~\cite{chen2023minigpt}&2023&\cellcolor{gray!30}\makecell[c]{SBU, LAION\\~~~~~~~~~~CC3M, GRIT-20M~~~~~~~~~~}&\cellcolor{gray!30}N/A&N&&&&&&&&\ding{51}&\\

SAM~\cite{kirillov2023segment}&2023&SA-1B&\makecell[c]{11 M images\\1.1 B masks}&N&&&\ding{51}&&&&&&			\\

\rowcolor{gray!30}SAT~\cite{zhao2023one}&2023&\cellcolor{gray!30}SAT-DS&22K volumes, 302K masks&N&&&&&&&&&\ding{51}			\\

Vita-CLIP~\cite{wasim2023vita}&2023&\makecell[c]{K400, SSV2}&N/A&N&&&&&\ding{51}&&&&\\	

\rowcolor{gray!30}CheXagent~\cite{chen2024chexagent}&2024&CheXinstruct&8.4 M image-text pairs&N&&&&&&&\ding{51}&&			\\		

Claude 3-Opus~\cite{anthropic2024claude}&2024&N/A&N/A&N&&&&&\ding{51}&&&&\ding{51}			\\		
\rowcolor{gray!30}Gemini 1.5 Pro~\cite{team2024gemini}&2024&N/A&N/A&N&&&&&&&&&\ding{51}			\\	

GPT-4V~\cite{openai2023gpt4v}&2024&N/A&N/A&N&\ding{51}&\ding{51}&\ding{51}&&\ding{51}&\ding{51}&\ding{51}&\ding{51}&\ding{51}		\\	

\rowcolor{gray!30}GPT-4 Turbo~\cite{achiam2023gpt}&2024&N/A&N/A&N&&&&&&&&&\ding{51}			\\

GPT-4o~\cite{hurst2024gpt}&2024&N/A&N/A&N&&&&&\ding{51}&&&&\ding{51}		\\

\rowcolor{gray!30}LLaMA 3~\cite{grattafiori2024llama}&2024&Self-Constructed&15T multilingual tokens&N&&&&&&&&\ding{51}&				\\

\hline

Clinical-BERT~\cite{yan2022clinical}&2022&MIMIC-CXR&227K image-text pairs&M&\ding{51}&  &  &  &\ding{51}&  &  &  &  \\

\rowcolor{gray!30}MedCLIP~\cite{wang2022medclip}&2022&CheXpert, MIMIC-CXR&\makecell[c]{223K images\\~~~~~~222K image-text pairs~~~~~~}&M&\ding{51}&  &  &  &  &  &  &  &\ding{51}			\\

BiomedCLIP~\cite{zhang2023biomedclip}&2023&PMC-15M&15 M image-text pairs&M	&\ding{51}&\ding{51}&\ding{51}&\ding{51}	&\ding{51}&\ding{51}		\\		

\rowcolor{gray!30}BioViL~\cite{bannur2023learning}&2023&MIMIC-CXR v2&179K image-text pairs&M&  &  &  &  &\ding{51}&  &  &  &  	\\				

CXR-CLIP~\cite{you2023cxr}&2023&\makecell[c]{CheXpert\\ChestX-ray14, MIMIC-CXR}&557K image-text pairs&M&  &  &  &  &\ding{51}&  &\ding{51}&  &  \\		

\rowcolor{gray!30}FLAIR~\cite{silva2025foundation}&2023&\cellcolor{gray!30}Self-Constructed&288K images, 101 text labels&M&  &  &  &  &\ding{51}&  &  &  & \\

KAD~\cite{zhang2023knowledge}&2023&MIMIC-CXR14&227K image-text pairs&M&  &  &\ding{51}&  &\ding{51}&  &  &  &  	\\	

\rowcolor{gray!30}LLaVA-Med~\cite{li2024llava}&2023&PMC-15M&660K image-text pairs&M&  &\ding{51}&  &  &  &  &  &  &  		\\			

MedKLIP\cite{wu2023medklip}&2023&MIMIC-CXR v2&227K image-text pairs&M&  &  &  &  &\ding{51}&  &  &  &  	\\			

\rowcolor{gray!30}M-flag~\cite{liu2023m}&2023&MIMIC-CXR&213K image-text pairs&M&  &  &\ding{51}&  &\ding{51}&  &  &  &\ding{51}		\\	

PMC-CLIP~\cite{lin2023pmc}&2023&PMC-OA&1.6M image-text pairs&M&  &  &  &\ding{51}&  &  &  &  &  			\\		

\rowcolor{gray!30}PubMedCLIP~\cite{eslami2023pubmedclip}&2023&ROCO&80K image-text pairs&M&  &\ding{51}&  &  &  &  &\ding{51}&  &  	\\		

Qilin-Med-VL~\cite{liu2023qilin}&2023&ChiMed-VL&\makecell[c]{580K image-text pairs\\469K question-answer pairs}&M&  &  &  &  &\ding{51}&  &  &  &  			\\		

\rowcolor{gray!30}Ret-CLIP~\cite{du2024ret}&2023&RET-Clinical&193K image-text pairs&M&  &  &  &  &\ding{51}&  &  &  &  \\		

BioMedLM~\cite{bolton2024biomedlm}&2024&PubMed&34.6B tokens&M&  &\ding{51}&  &  &  &  &  &  &  				\\	

\rowcolor{gray!30}CT-CLIP~\cite{hamamci2024developing}&2024&CT-RATE&50K volumes-text pairs&M&  &\ding{51}&  &  &  &  &  &  &  	\\			

Maco~\cite{huang2024enhancing}&2024&MIMIC-CXR v2&227K image-text pairs&M&  &  &\ding{51}&  &\ding{51}&  &  &  &\ding{51}		\\	

\rowcolor{gray!30}Medical X-VL~\cite{park2024self}&2024&MIMIC-CXR&227K image-text pairs&M&\ding{51}&  &  &  &  &  &  &  &  	\\

MPMA~\cite{zhang2023multi}&2024&ROCO, MIMIC-CXR&235K image-text pairs&M&\ding{51}&\ding{51}&  &  &\ding{51}&  &  &  &  				\\

\rowcolor{gray!30}QFT~\cite{su2024design}&2024&\cellcolor{gray!30}Self-Constructed&\cellcolor{gray!30}\makecell[c]{10,720 images\\5,360 Chinese reports\\~~23K question-answer pairs~~~}&M&\ding{51}&  &\ding{51}&  &\ding{51}&  &  &  &\ding{51}		\\

Unichest~\cite{dai2024unichest}&2024&Multi-CXR&\makecell[c]{685K images\\348K image-text pairs}&M&  &  &  &  &\ding{51}&  &  &  &\ding{51}					\\

\rowcolor{gray!30}UniMed-CLIP~\cite{khattak2024unimed}&2024&UniMed&5.3M image-text pairs&M&  &  &  &  &\ding{51}&  &  &  &  	\\

\bottomrule
\end{tabular}}
\end{table*}

\textbf{Improving Explainability and Medical Knowledge Integration}.
Given the high-stakes nature of medical imaging, explainable AI (XAI) is essential. Several studies have proposed explainable prompt learning frameworks to enhance interpretability. XCoop~\cite{bie2024xcoop} introduces clinically guided concept-driven prompts to align the semantics of medical images with textual descriptions, offering both visual and textual explanations. Additionally, knowledge graph embeddings for radiology reports~\cite{van2023knowledge} leverage multi-lingual structured medical knowledge (e.g., SNOMED CT) to improve disease and image classification without requiring extensive pretraining, thereby enhancing interpretability and efficiency. These approaches highlight the importance of integrating external medical knowledge into VLMs to improve clinical applicability.

\textbf{Addressing Low-Resource Challenges with Automated Prompt Generation}.
To mitigate the challenges of limited labeled data in medical imaging, researchers have explored automatic prompt generation methods. MedPrompt~\cite{zheng2024exploring} leverages unsupervised pretraining on large-scale medical images and texts to generate prompts without manual annotation, achieving superior zero-shot and few-shot performance compared to handcrafted prompts. Similarly, a novel low-shot prompt tuning framework for multiple-instance learning (MIL) in histopathology~\cite{chikontwe2024low} integrates residual visual feature adaptation and context-aware prompt optimization, significantly enhancing whole slide image (WSI) classification. In genetic biomarker prediction~\cite{zhang2024prompting}, large language models (LLMs) are employed to generate medical prompts, guiding the extraction of pathological instances and interactions, achieving an AUC of 91.49 \% in microsatellite instability (MSI) classification. These studies demonstrate the potential of prompt learning in reducing dependency on expert-designed prompts and improving model adaptability in low-resource settings.

\begin{figure*}[!t]
\centering
\includegraphics[width=\textwidth]{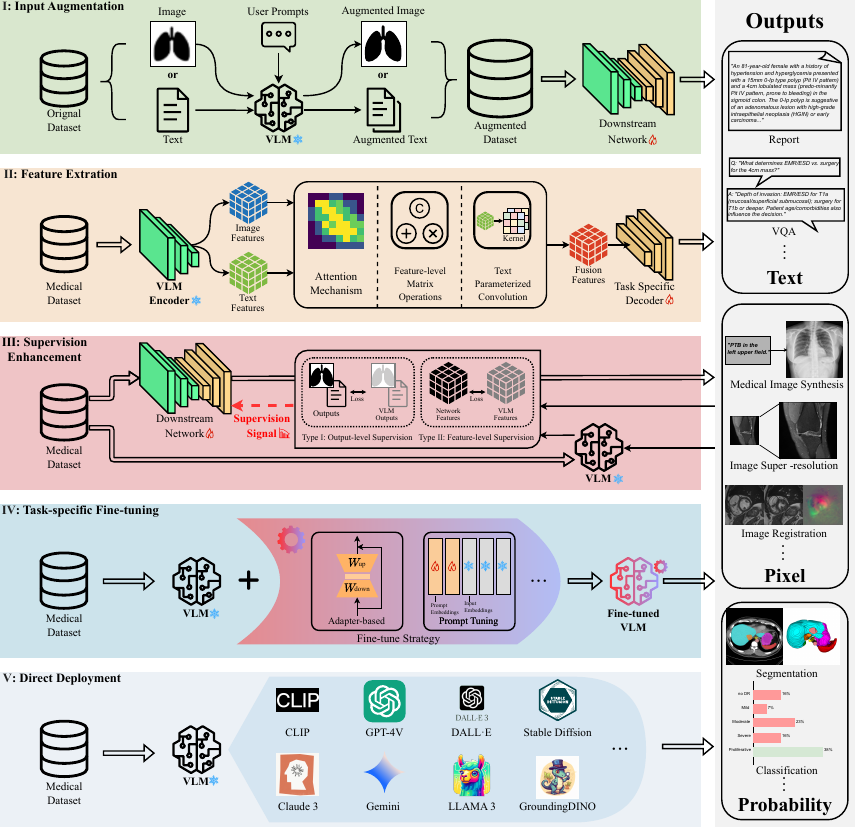}
\caption{Illustration of five VLM adaptation strategies to medical applications.}
\label{Fig2}
\end{figure*}

\section{Adaptation Strategies for VLMs in Medical Image Analysis Tasks}
\label{sec:Adaptation strategis}

Vision-Language Models (VLMs) have demonstrated remarkable capabilities in comprehending both textual and visual modalities, thereby enabling their utilization in a wide spectrum of multi-modal medical image analysis applications. Nevertheless, the unique characteristics of medical data, marked by high domain specificity and an intrinsic imbalance across modalities, pose significant challenges to the effective deployment of such models. To mitigate these issues, various adaptation strategies have been proposed to tailor VLMs for medical image analysis tasks. We categorize the existing methodologies into five principal strategies: input augmentation, feature extraction, supervision enhancement, task-specific fine-tuning, and direct deployment. The conceptual framework for these five strategies is summarized in Fig.~\ref{Fig2}, illustrating their respective processing pipelines. We discuss each strategy in detail in the following section.

\subsection{Input Augmentation}
Due to concerns regarding patient privacy, high annotation costs, and the heterogeneity of imaging protocols across institutions, existing medical datasets are often constrained by limited size, low diversity, and modality imbalance. These limitations are frequently compounded by imaging artifacts such as blur and noise, which further hinder model training and generalization, especially in tasks requiring fine-grained visual understanding.

Recent advances have explored leveraging VLMs to enhance and augment such datasets by capitalizing on their cross-modal capabilities. These approaches can be broadly categorized into the following two strategies:

\textbf{Image Generation}. Several studies have utilized VLMs to synthesize diverse training samples, thereby increasing the scale and heterogeneity of medical datasets while mitigating overfitting. Akrout et al.~\cite{akrout2023diffusion}, for instance, integrated diffusion models with fine-grained textual prompts to produce high-fidelity synthetic images for skin disease classification. Their findings indicate that training on entirely synthetic datasets significantly improves both data diversity and classification accuracy.

\textbf{Text Generation and Refinement}.
Medical models relying solely on imaging modalities often struggle to capture high-level semantic information critical for accurate diagnosis. Such information can be more effectively conveyed and supplemented through descriptive text. However, the scarcity of textual annotations in most medical datasets presents a barrier to integrating semantic cues into the diagnostic workflow.
To address this issue, VLMs have been employed to generate semantically rich textual descriptions directly from medical images. These descriptions serve to inject prior domain knowledge and contextual understanding into the visual analysis pipeline. 

For instance, Chowdhury et al.~\cite{chowdhury2024adacbm} proposed AdaCBM, an adaptive concept bottleneck model that leverages GPT-4 to generate 760 clinically relevant concepts for the HAM10000 dataset~\cite{tschandl2018ham10000}, thereby addressing the lack of textual annotations. Similarly, TV-SAM~\cite{jiang2024tv} employs medical concept descriptions generated by GPT-4 in response to clinician queries and medical images, using them as textual prompts to enhance zero-shot segmentation performance and facilitate improved clinician interaction for diagnostic purposes.

In other scenarios where textual annotations are available, they are often limited to brief, high-level descriptors (e.g., organ names) and lack crucial details pertaining to imaging characteristics, pathological findings, and other diagnostic indicators. To mitigate this limitation, several studies have explored the augmentation of existing annotations using VLMs to produce more comprehensive and semantically rich descriptions. In the domain of medical image synthesis, Chen et al.~\cite{chen2024medical} introduced an anatomy-pathology prompting framework that constructs anatomical and pathological vocabularies derived from expert-curated radiology reports. This framework then employs GPT-4 to generate detailed, descriptive reports that provide essential semantic context for synthesizing medically plausible images with high anatomical and pathological fidelity.

Furthermore, to support broader accessibility and facilitate cross-lingual research, the BIMCV-R dataset~\cite{chen2024bimcv} incorporates a GPT-4-based translation pipeline that converts Spanish radiology reports into English, followed by expert-level human verification. This process effectively eliminates language barriers, ensures the standardization of medical reports, and enhances their usability in downstream cross-modal retrieval and diagnostic tasks.

\subsection{Feature Extraction}
Beyond input augmentation, numerous methods leverage the pre-trained encoders of VLMs to extract deep feature representations from medical data. These encoders, trained on large-scale image-text datasets, demonstrate strong capabilities in capturing high-level semantic information from both visual and textual modalities. Specifically, they can accurately identify morphological characteristics of lesion regions and extract contextual semantics that are well-aligned with clinical narratives.

In clinical settings, accurate and robust diagnosis often requires the integration of heterogeneous data from multiple modalities. To this end, a growing body of research has explored cross-modal fusion strategies that combine VLM-extracted visual features with clinical text embeddings. This fusion enhances modality complementarity and inter-modal interactions, ultimately improving task-specific performance across diverse medical applications.

To facilitate this integration, task-specific decoders are commonly employed to project the fused representations into output spaces tailored to distinct clinical objectives. Existing fusion strategies typically follow one of the following methodological frameworks:

\textbf{Attention mechanism}. Attention mechanisms enable neural networks to dynamically prioritize and focus on the most salient components of input data. Among them, cross-attention mechanisms compute attention scores between queries from one modality and key-value pairs from another, thereby facilitating explicit interactions between heterogeneous representations. These mechanisms are particularly effective in medical multi-modal tasks, where alignment between visual and textual data is critical.

Bai et al.~\cite{bai2023cat} introduced a Guided Attention Module based on cross-attention to align visual and textual features for Visual Question Localized Answering (VQLA) in robotic surgery. The cross-attention mechanism ensures precise semantic alignment by correlating relevant image regions with textual queries, effectively mitigating multi-modal ambiguity and enabling accurate answer generation.

Similarly, Feng~\cite{feng2024enhancing} employed cross-attention to align visual features derived from a denoising diffusion probabilistic model (DPM) ~\cite{ho2020denoising} with textual diagnostic annotations. This alignment enhances the semantic richness of DPM-generated visual representations, facilitating more effective segmentation. By integrating cost-efficient textual data with semantically dense image features, this approach significantly reduces reliance on labor-intensive manual annotations while maintaining high segmentation performance.

Some works combine cross-attention with other strategies to meet task requirement.  
Zhao et al. \cite{zhao2024cap2seg} leverage the scale-aware textual attention module, a self-attention mechanism, refines hierarchical and multi-scale textual features, enhancing the cross-attention in the language-aware visual decoder for precise multi-modal alignment, and addressing the challenges of scale variance in segmentation tasks.
Liu et al. \cite{liu2024deep} combine cross-attention with a multiteacher knowledge distillation (MKD) module to tackle the inherent uncertainty and vagueness often faced by traditional fusion modules in multi-modal VQA tasks.
The MKD module enhances the representation of incomplete modalities through features generated from the complete modality. 
This synergistic approach improves semantic interactions and enhances robustness against common noise in medical data.

In addition to cross-attention, various other attention mechanisms have been adopted for multi-modal feature fusion in medical applications. Zhan et al.~\cite{zhan2023debiasing} employed Bilinear Attention Networks (BAN)~\cite{kim2018bilinear} to integrate multi-modal features for VQA. Unlike cross-attention, which focuses on direct query-to-feature associations, bilinear attention facilitates richer pairwise feature interactions, thereby enhancing the model’s ability to generate more accurate and interpretable results.

Furthermore, the Gated Attention Mechanism (GAM) proposed in \cite{song2022multimodal} adaptively modulates the contribution of each modality, thereby strengthening the complementary relationships among three modalities and improving overall classification performance. GAM prioritizes the primary modality (e.g., pathological images) while dynamically adjusting the influence of auxiliary modalities, leading to improved robustness and interpretability in clinical contexts.

To further enhance visual feature fusion, the Pyramid Squeeze Attention (PSA) mechanism~\cite{zhang2022epsanet} integrates spatial and channel attention strategies as introduced in \cite{li2024efficient} to capture multi-scale information. By simultaneously focusing on local visual details and global contextual cues, PSA significantly boosts the network’s discriminative capability, particularly in challenging tasks such as medical text detection involving multi-oriented and densely packed text instances.

\textbf{Matrix operation}.
In contrast to attention-based mechanisms that involve complex computational processes, simple matrix operations, such as addition, multiplication, and concatenation, provide an efficient alternative for multi-modal integration with minimal computational overhead. These operations are highly adaptable and can be tailored to meet diverse task requirements while preserving efficiency. Among them, concatenation is the most widely adopted strategy.

For example, the PMFCC module in fTSPL~\cite{wang2024ftspl} calculates Pearson correlations between text embeddings and voxel-level image features, then concatenates them with functional connectivity data to construct a comprehensive multi-modal representation. This integration enhances both brain disease classification and cognitive function prediction. Similarly, Chen et al.~\cite{chen2024novel} proposed MM-UniCMBs, a framework that fuses text descriptions with 3D MRI features via dimensional transformation and concatenation. A transformer-based fusion module is then applied, leveraging clinical knowledge to improve the detection and classification of cerebral microbleeds.

Element-wise multiplication also plays a crucial role in multi-modal fusion. Yu et al.~\cite{yu2024language} introduced a local processing branch that combines organ-level text embeddings with image-derived features through element-wise multiplication for multi-organ trauma detection. Likewise, the Voxel-Wise Prompts Module in VCLIPSeg~\cite{li2024vclipseg} fuses CLIP-based text embeddings with voxel-level image features via element-wise multiplication to create a shared semantic space, which facilitating robust and generalized semi-supervised segmentation under limited annotation conditions.

Matrix addition is another effective strategy. In LViT~\cite{li2023lvit}, text embeddings are added to image features at the pixel level within a U-shaped Vision Transformer (ViT) branch, resulting in a unified multi-modal representation. This design enhances segmentation performance, particularly in semi-supervised scenarios.

\textbf{Text-parameterized convolution}.
The Text-Parameterized Convolution (TPC) operation employs textual representations to dynamically generate the weight ($W$) and bias ($b$) parameters of a convolutional layer, which are subsequently applied to visual features via convolution operations. This method enables the model to adapt visual feature extraction dynamically based on semantic cues from the text, thereby achieving more fine-grained cross-modal integration. For example, Chen et al.~\cite{chen2024causalclipseg} project the text features into a $D$-dimensional embedding space using a learnable linear projection, where $D = C \times K \times K + 1$. The resulting embedding is then decomposed into $W \in \mathbb{R}^{C \times K \times K}$ and $b \in \mathbb{R}$, which are used as the convolutional layer's parameters and applied to the visual input. Given that convolutional layers inherently model spatial hierarchies, this approach allows the textual modality to influence the visual feature extraction process across multiple levels of abstraction, thereby improving the localization of lesions of varying sizes in medical image segmentation tasks.

\textbf{Attaching task-specific decoder}.
As the clinical objectives of many downstream applications differ significantly from those addressed during VLM pre-training, it is essential to incorporate task-specific decoders for different applications.

Silva et al.~\cite{silva2023exploring} proposed the use of specialized classifiers that leverage FLAIR’s pre-trained features to improve diagnostic performance in fundus imaging tasks, including the detection of hypertensive retinopathy. Similarly, Müller et al.~\cite{muller2025chex} integrated multitask output heads, including bounding box prediction modules and regional explanation generators, into VLMs for chest X-ray analysis. These additions enable interactive and interpretable localization and textual description of radiological findings.

\subsection{Supervision Enhancement}
Recent studies have attempted to incorporate VLMs into the loss computation process for medical analysis tasks. By leveraging the rich prior knowledge embedded in VLMs from large-scale pre-training, these approaches enhance the optimization of downstream clinical objectives.

Based on a comprehensive review of the literature, these methods can be broadly categorized into two main paradigms: output-level supervision and feature-level supervision. Each approach addresses distinct challenges in model training and generalization through tailored strategies, as described in the following subsections.

\textbf{Output-level supervision}. Training a reliable network typically relies on access to large-scale, high-quality annotated datasets. However, acquiring extensively fine-grained annotations in the medical domain presents substantial challenges. On one hand, such annotations demand domain-specific expertise, as only qualified medical professionals can provide accurate and clinically reliable labels. On the other hand, the annotation process is prohibitively time-consuming and labor-intensive, particularly for fine-grained segmentation tasks requiring pixel-level delineation of anatomical or pathological structures. To mitigate these challenges, recent studies have explored using VLMs to automatically generate structured image-text pairs or pseudo-labels that approximate expert-level annotations, thereby expanding training data while reducing reliance on manual labeling.

Lanfredi et al.~\cite{lanfredi2025enhancing} employed a locally executable VLM to extract structured annotations, such as presence, probability, severity, and anatomical location of abnormalities from chest X-ray reports. These multi-attribute labels are then used to train a classification model for chest X-ray abnormality detection. The resulting annotations provide fine-grained and diverse supervision signals, enabling models to focus on clinically relevant regions and improving both diagnostic accuracy and robustness.

Koleilat et al.~\cite{koleilat2024medclip} proposed MedCLIP-SAM, a pipeline that leverages gScoreCAM to generate saliency maps from BiomedCLIP, which are used to create initial segmentation masks. These masks are subsequently refined into bounding-box prompts for the Segment Anything Model (SAM), resulting in high-quality pseudo-masks. These pseudo-labels serve as supervision in a weakly supervised learning framework, allowing for improved segmentation accuracy while minimizing the need for costly manual annotations.

\textbf{Feature-level supervision}. 
The inherent modality gap between medical texts and images presents a fundamental challenge for effective cross-modal feature alignment. Misaligned feature fusion often introduces semantic noise and ambiguities, thereby compromising inter-modal consistency and leading to suboptimal or inaccurate predictions.

While output-level loss functions can guide global optimization, they lack the capacity to explicitly model fine-grained alignment relationships between modalities. Consequently, such loss functions are limited in their ability to correct local errors arising from semantic mismatches. To address this limitation, recent studies have leveraged VLMs to extract both visual and textual features, and designed loss functions based on multi-modal similarity measures. This approach provides more direct and localized supervisory signals, facilitating precise alignment between fine-grained visual concepts and detailed textual descriptions.

For instance, Zhu et al.~\cite{zhu2024stealing} introduced a contrastive loss function to align text embeddings from pre-trained VLMs on the server side with image features from client devices, addressing classifier bias in federated learning under heterogeneous medical data distributions. Similarly, Bai et al.~\cite{bai2025surgical} employed contrastive training with adversarial examples in their Surgical-VQLA++ framework to improve robustness and generalization in visual question-localized answering (VQLA) for robotic surgery. This method utilizes the pre-trained DeiT model~\cite{touvron2021training} to jointly process visual and textual inputs, while incorporating adversarial perturbations. The model is trained to differentiate clean from corrupted embeddings, enhancing resilience against noise, artifacts, and imaging distortions, factors critical in real-world surgical scenarios.

CLIP-Lung~\cite{lei2023clip} introduced three knowledge-guided contrastive loss functions to align clinical text annotations with image features. These losses include: image-class alignment to improve classification accuracy; image-attribute alignment to capture subtle visual details; and class-attribute alignment to strengthen the understanding of semantic relationships in textual inputs. All image features and text embeddings used for loss computation are encoded by VLMs. Collectively, these strategies enable the model to better distinguish visually similar lung nodules, predict malignancy progression, and offer more interpretable diagnostic insights.

\subsection{Task-Specific Fine-Tuning}
While many widely adopted VLMs are pre-trained on large-scale natural image datasets such as ImageNet~\cite{deng2009imagenet} and COCO~\cite{lin2014microsoft}, their direct application to medical imaging often yields suboptimal performance due to substantial domain shifts in data distribution and semantic context.

To bridge this domain gap, transfer learning has emerged as a pivotal strategy, wherein networks are initialized with pre-trained VLM weights and subsequently fine-tuned using domain-specific medical datasets. Recent approaches retain the core VLMs as fixed backbones while introducing lightweight, task-specific modules (e.g., adapter layers or prompt-tuning components) for efficient adaptation. This fine-tuning paradigm strikes a favorable balance between preserving generalizable vision-language representations and adapting to domain-specific semantics. Moreover, it requires only limited annotated data and moderate computational resources, making it particularly practical for real-world clinical deployment, where data scarcity and infrastructure constraints are common.
Current task-specific fine-tuning strategies encompass the following representative genres:

\textbf{Prompt tuning}.
Prompt tuning is a parameter-efficient technique that incorporates learnable embeddings into the input space of pre-trained VLMs, optimizing only these prompts during fine-tuning. Jia et al.~\cite{jia2022visual} formalized this approach to adapt general VLMs for downstream tasks. CITE~\cite{zhang2023text} applies prompt tuning by introducing a small number of trainable parameters that bridge biomedical text embeddings and visual encoders, thereby enhancing pathological image classification under low-data regimes. Similarly, PPAD~\cite{sun2024position} introduces position-guided prompt tuning using learnable text and image prompts to adapt CLIP models for chest X-ray anomaly detection. This strategy mitigates domain discrepancies and improves localization of critical lung regions for enhanced diagnostic accuracy.

\textbf{Adapter-based tuning}. Adapter modules function similarly to prompt tuning but adopt a modular architecture. A typical adapter comprises a down-sampling linear layer, a nonlinear activation, normalization, an up-sampling linear layer, and a residual connection~\cite{houlsby2019parameter}. These lightweight components are inserted into existing model layers to enable efficient fine-tuning. Ding et al.~\cite{ding2024hia} proposed the High-resolution Instruction-Aware Adapter (HiA), designed to enhance Chinese medical multi-lingual VLMs by supporting high-resolution image input and integrating instruction-specific cues. HiA addresses challenges such as small lesion detection and temporal image sequence understanding. MAIRA-2~\cite{bannur2024maira} introduces a chest X-ray-specific multi-modal architecture featuring novel adapter designs that incorporate spatial grounding of clinical findings and comprehensive contextual inputs. This design improves the accuracy, completeness, and verifiability of radiology report generation.

\textbf{Other fine-tuning strategies}.
In certain clinical tasks, conventional fine-tuning approaches are insufficient due to data scarcity and task complexity. To overcome this, Van et al.~\cite{van2023open} applied prefix tuning to fine-tune GPT2-XL~\cite{radford2019language} for medical VQA. Unlike prompt tuning, prefix tuning introduces learnable embeddings across multiple transformer layers, enabling deeper interaction with internal attention mechanisms. This approach improves the generation of context-aware, accurate free-form answers in specialized, low-resource medical domains.

\subsection{Direct Deployment}
Practical clinical scenarios often involve labor-intensive tasks, highlighting the need for automated systems to alleviate clinician workload and allow greater focus on critical decision-making. To address this, several studies have investigated the direct application of VLMs to downstream tasks involving complex multi-modal medical data, aiming to assess their diagnostic capabilities without requiring additional model training.

This category encompasses use cases such as medical image generation, disease diagnosis, and VQA, wherein the output format remains consistent with general-purpose VLMs. For example, Temsah et al.~\cite{temsah2024art} employed DALL·E 3~\cite{betker2023improving} to generate synthetic images of congenital heart disease for medical education. While the study revealed some educational value, it also identified substantial limitations in clinical accuracy and representational fidelity. Upadhyaya et al.~\cite{upadhyaya2024360} proposed a multi-view prompting framework that incorporates feedback from medical experts to guide Gemini~\cite{team2023gemini} in diagnosing amblyopia using eye-tracking data. Without fine-tuning, the system demonstrated significant improvements in classifying amblyopia severity, subtypes, and associated conditions such as nystagmus.

Yeh et al.~\cite{yeh2024insight} introduced a multi-modal diagnostic pipeline that integrates an instance segmentation network as a visual translator to process meibography images. This module extracts quantitative morphological data, such as gland atrophy, density, and tortuosity, which are then embedded into VLM prompts along with clinical metadata to produce interpretable diagnostic outputs. This pipeline enables detailed morphological analysis, enhancing the clinical relevance and interpretability of VLM-generated results.

Other research efforts have proposed novel frameworks and strategies to adapt existing VLMs to specific clinical workflows. These methods operate in a zero-shot or direct deployment setting, leveraging the VLM’s pre-trained capabilities without further fine-tuning. This makes them particularly appealing for rapid deployment in healthcare environments where annotated data and computational resources are limited. Xplainer~\cite{pellegrini2023xplainer} exemplifies the zero-shot capabilities of VLMs in clinical contexts. It is a novel framework that classifies descriptive observations in chest X-rays without labeled training data. By leveraging VLMs for multi-label classification, Xplainer delivers both accurate and explainable diagnostic predictions, offering a promising path toward interpretable, data-efficient medical AI.

\begin{figure*}[!t]
\centering
\includegraphics[width=\textwidth]{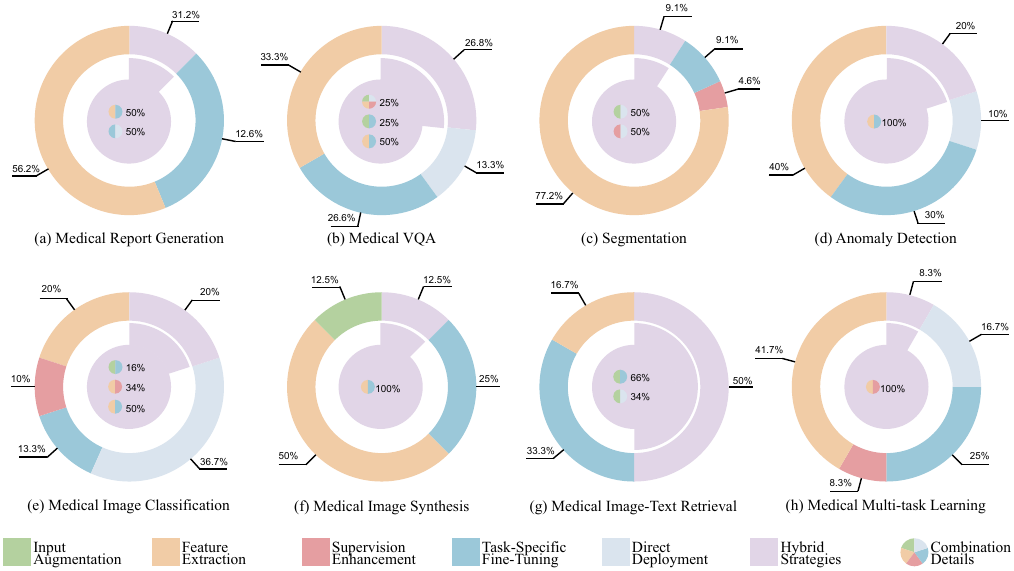}
\caption{Overview of how different VLM adaptation strategies are allocated among diverse medical tasks in the analyzed literature. Hybrid strategies refer to combinations composed of two or more individual strategies.}
\label{Fig3}
\end{figure*}

\begin{figure*}[!t]
\centering
\includegraphics[width=\textwidth]{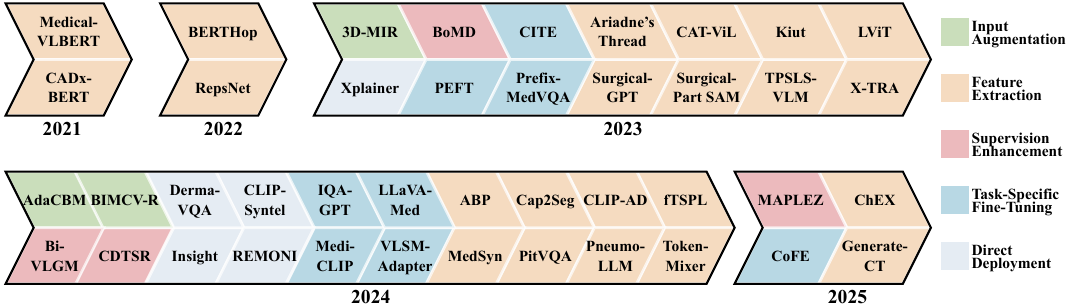}
\caption{The chronological evolution of VLMs adaptation strategies in medical, along with representative studies corresponding to each category.}
\label{Fig4}
\end{figure*}

\section{Taming VLMs for Medical Image Analysis Tasks}
To enable the practical deployment of Vision-Language Models (VLMs) in clinical settings, this section explores five representative adaptation strategies across core medical applications, including medical report generation, visual question answering (VQA), image segmentation, anomaly detection, image classification, image synthesis, image-text retrieval, and multi-task learning. Fig.~\ref{Fig3} presents the distribution of adaptation strategies across various tasks, while Fig.~\ref{Fig4} illustrates their chronological development, highlighting representative approaches and emerging trends in medical VLM research.

\subsection{Medical Report Generation}
Medical report generation aims to automatically produce descriptive, clinically coherent diagnostic narratives based on medical images. This process assists radiologists by reducing documentation workload and improving diagnostic interpretability. Formally, the task can be defined as: $S^* = \arg\max_{S} P(S | I; \theta)$, where $P$ is the conditional probability of generating a report $S$ given medical image $I$, and $\theta$ represents the model parameters. 
The objective is to obtain $S^*$, the most probable and clinically relevant report corresponding to $I$.
Conventional deep learning-based approaches to this task face two major limitations: 1) Low Text Generation Quality: Generated reports often exhibit templated structure, weak semantic coherence, and inadequate logical relationships between clinical terms. 2) Multi-Modal Fragmentation: These models typically depend on unimodal processing pipelines, resulting in limited interaction between visual and textual modalities.

The integration of VLMs offers promising advancements in overcoming these challenges: 1) Incorporation of Language Model Priors: Leveraging large-scale pre-trained language models (e.g., LLaMA infused with medical domain knowledge) and reinforcement learning techniques significantly enhances the semantic coherence and fluency of generated reports. 2) Multi-Modal Fusion: VLM architectures enable dynamic and fine-grained integration of image and text features, improving inter-modal synergy and enabling more context-aware and accurate generation. By incorporating VLMs into medical report generation workflows, it is possible to improve diagnostic quality, reduce clinician workload, and enhance the overall interpretability and efficiency of radiological reporting systems.

\subsubsection{Adapting VLMs to Medical Report Generation}
\textbf{Feature Extraction}. Various studies have investigated different multi-modal fusion strategies utilizing VLMs, each contributing to enhanced accuracy, semantic coherence, and clinical relevance in automated medical report generation.

One commonly adopted approach is concatenation-based feature fusion, which combines image and text embeddings to form a unified representation. Kapadnis et al.~\cite{kapadnis2024serpent} introduced SERPENT-VLM, which employs a concatenation strategy to integrate image and text features while refining their alignment to reduce hallucinations in radiology reports. Huang et al.~\cite{huang2023kiut} utilized concatenation within a knowledge-injected U-Transformer architecture to strengthen multi-modal feature integration. Similarly, Liu et al.~\cite{liu2021medical} and Fonolla et al.~\cite{fonolla2021automatic} applied concatenation to merge visual and textual representations, demonstrating that such integration effectively enhances report generation performance.

Attention-based fusion mechanisms have also been widely employed to facilitate deeper cross-modal interactions. Tanwani et al.~\cite{tanwani2022repsnet} implemented attention-driven fusion within an encoder-decoder architecture to align visual and textual features. Xu et al.~\cite{xu2023vision} proposed a vision-knowledge fusion framework that incorporates attention over knowledge graphs, enhancing interpretability and semantic richness in generated reports. More specialized techniques have also emerged. Huh et al.~\cite{huh2023improving} introduced an X-shaped cross-attention mechanism that alternates between image-to-text and text-to-image fusion to align modalities within a shared space, aiding efficient text correction training. Tan et al.~\cite{tan2024medical} employed multi-headed attention to fuse visual and audio-textual features, generating reports grounded on image regions highlighted by auxiliary signals, thereby mitigating the issue of imbalanced data distribution commonly found in medical datasets.

Departing from traditional concatenation and attention paradigms, Yang et al.~\cite{yang2024token} introduced Token-Mixer, a dynamic module that intermixes image and text tokens within a shared embedding space. This approach mitigates exposure bias in auto-regressive text generation, enhancing cross-modal alignment and improving the overall quality of generated reports.

In summary, the integration of VLMs into medical report generation has been advanced through diverse fusion strategies. These works collectively underscore the critical role of effective multi-modal feature fusion in producing coherent, accurate, and clinically meaningful diagnostic narratives.




\textbf{Task-Specific Fine-Tuning}.
Fine-tuning VLMs for medical report generation plays a crucial role in enhancing domain alignment and improving the factual accuracy and clinical relevance of generated reports.

Liu et al.~\cite{liu2024bootstrapping} fine-tuned VLMs for radiology report generation using a contrastive learning objective to improve domain alignment, alongside a coarse-to-fine decoding strategy for generating more refined and coherent textual outputs. Similarly, Li et al.~\cite{li2025contrastive} applied fine-tuning with counterfactual explanations to mitigate biases and improve the factual consistency of generated radiology reports.

Extending beyond radiology, Tan et al.~\cite{tan2024clinical} proposed a multi-scale vision transformer (MR-ViT) fine-tuned for pathology report generation. This model aligns whole-slide images (WSIs) with real pathology reports to ensure clinical fidelity. Zhou et al.~\cite{zhou2024pre} fine-tuned SkinGPT-4, a dermatology-focused VLM that integrates a pre-trained vision transformer with LLaMA-2, enabling diagnostic and treatment recommendation generation in dermatological scenarios.

Another promising direction is retrieval-augmented fine-tuning. Jeong et al.~\cite{jeong2024multimodal} fine-tuned X-REM, a retrieval-enhanced model that optimizes image-text alignment for report generation through multi-modal retrieval mechanisms.

Collectively, these studies demonstrate the effectiveness of domain-specific fine-tuning strategies in adapting general-purpose VLMs for medical report generation across diverse clinical domains, including radiology, pathology, and dermatology.


\textbf{Hybrid strategies}. Recent advancements in medical report generation have explored multi-category approaches by integrating multiple VLM strategies, aiming to enhance both task performance and model adaptability across diverse clinical scenarios.

Sun et al.~\cite{sun2024continually} proposed a hybrid framework that combines concatenation-based fusion of textual and visual features to enhance cross-modal representation. Additionally, they incorporate contrastive learning for improved alignment between modalities. To enable effective knowledge transfer from X-ray to CT imaging, a 2D–3D adapter module is introduced. This component facilitates modality adaptation while keeping the majority of Large Language Model (LLM) parameters frozen, thereby preserving generalization capabilities. This approach exemplifies a combination of Supervision Enhancement and Task-specific Fine-tuning.

Chen et al.~\cite{chen2024iqagpt} proposed a hybrid VLM–LLM pipeline that fine-tunes a captioning VLM to generate descriptive quality reports. These reports are subsequently processed by ChatGPT for quality scoring and radiological summarization, demonstrating the synergistic potential of VLMs and LLMs in clinical applications. This method combines elements from \textbf{Category \Rmnum{4}} (Fine-Tuning) and \textbf{Category \Rmnum{5}} (Direct Deployment).

These studies collectively underscore the potential of integrating multiple VLM strategies to improve the adaptability and scalability of automated medical report generation systems, particularly when applied across heterogeneous imaging modalities and clinical tasks.



\subsubsection{Discussion}
While VLMs demonstrate significant potential for automated medical report generation, several critical challenges limit their practical adoption in clinical environments.

One prominent issue is localized perception deficiency, wherein VLMs struggle to detect subtle pathological features, such as microcalcifications or early-stage lesions, with sufficient accuracy. This limitation often results in incomplete documentation of abnormalities, primarily due to inadequate alignment between cross-modal feature representations and the nuanced requirements of radiological detection. To mitigate this, multi-phase training strategies can be employed to enhance spatial-semantic consistency while maintaining diagnostic fidelity.

Another key challenge is structural discontinuity in generated reports. VLMs frequently produce outputs that fail to adhere to established clinical documentation standards, leading to illogical transitions between report sections and the omission of critical elements, such as differential diagnoses or follow-up recommendations. This lack of structural and semantic coherence diminishes the utility of VLM-generated content in real-world diagnostic workflows. One promising direction is the integration of clinical-pathway-constrained evaluation metrics, which assess linguistic quality based on adherence to structured medical reasoning and reporting protocols.

Future research should prioritize enhancing anatomical precision and clinical reasoning in report generation by incorporating explicit region-level localization and standardized reporting formats. The development of unified frameworks that jointly optimize spatial accuracy and logical narrative coherence will be essential for aligning VLM outputs with real-world clinical expectations and workflows.

\subsection{Medical VQA}
 
The primary objective of Medical Visual Question Answering (VQA) is to develop AI systems capable of interpreting medical images and answering clinically relevant natural language questions. Such systems aim to support diagnostic decision-making, enhance clinician efficiency, and promote patient engagement. Formally, the task can be defined as: $A=f(I,Q)$, where $A$ is the predicted answer, $I$ is the medical image, $Q$ is the input question, and $f$ denotes the model mapping multi-modal inputs to answers.
Medical VQA tasks are generally categorized into two formats: closed-ended and open-ended. In the closed-ended setting, a finite set of predefined answer choices is provided, allowing the problem to be formulated as a classification task. In contrast, open-ended VQA requires the model to generate free-form answers, necessitating a deeper understanding of medical visual content and the ability to process potentially ambiguous or patient-generated queries.

Traditional deep learning-based Medical VQA models encounter significant limitations. Data scarcity and annotation imbalance impair model generalization, while the lack of external medical knowledge integration restricts the accuracy and interpretability of the generated responses.

The emergence of VLMs offers a promising solution to these challenges. Their few-shot and self-supervised learning capabilities enable better utilization of large-scale pre-trained models and allow efficient adaptation to smaller, domain-specific datasets. Furthermore, knowledge-enhanced architectures, such as retrieval-augmented transformers and prompt engineering with embedded medical knowledge, facilitate seamless integration of external information sources, improving both response accuracy and clinical relevance.

Incorporating VLMs into Medical VQA pipelines significantly enhances performance by improving cross-domain generalization, reducing clinician workload in real-world scenarios, and strengthening the feasibility of deploying AI-powered assistants in clinical medical imaging workflows.

\subsubsection{Adapting VLMs to Medical VQA}
Recent advances in Medical VQA have demonstrated diverse strategies for incorporating VLMs, with notable contributions spanning across feature extraction, task-specific fine-tuning, direct deployment, and various hybrid-strategy approaches.

\textbf{Feature Extraction}. Several Medical VQA methods employ VLMs to facilitate deep cross-modal fusion, particularly through attention mechanisms and matrix operations. He et al.~\cite{he2024pitvqa} utilized cross-attention-based text encoders to inject visual features into textual representations, thereby grounding semantic concepts within surgical imagery and improving comprehension. Similarly, Naseem et al.~\cite{naseem2022vision} alternately set image and question features as query, key, and value components within the Transformer’s attention block, allowing for effective and flexible multi-modal integration. 

Gong et al.~\cite{gong2022vqamix} introduced VQAMix, a novel data augmentation framework that significantly improves performance across attention-based models such as BAN~\cite{kim2018bilinear} and SAN~\cite{yang2016stacked}, both of which utilize attention-driven fusion. Additional approaches, including Zhan et al.~\cite{zhan2023debiasing} and Liu et al.~\cite{liu2024deep}, also fall under this category, as discussed previously. 

For matrix-based fusion, Seenivasan et al.~\cite{seenivasan2023surgicalgpt} adopted a token ordering strategy by concatenating word tokens before vision tokens. This ensures that the model comprehends the question context prior to processing image features, thus improving answer inference quality.

\textbf{Task-Specific Fine-Tuning}. VLMs are inherently capable of generating answers from paired image-question inputs, making them naturally suitable for Medical VQA. Domain-specific fine-tuning allows these models to adapt effectively to clinical use cases.

LLaVA-Med~\cite{li2024llava} extends LLaVA for biomedical applications using a two-stage curriculum learning pipeline. The first stage establishes concept grounding by aligning biomedical vocabulary through figure-caption pairs. The second stage leverages GPT-4–generated instruction-following data to enhance open-ended reasoning and conversational skills for accurate biomedical VQA.

Van et al.~\cite{van2023open} integrated CLIP-based visual features into a pre-trained language model by mapping them into learnable tokens, which act as a visual prefix. To minimize computational overhead while preserving performance, the framework also incorporates LoRA and prompt tuning techniques.

\textbf{Direct Deployment}. Some studies evaluate the performance of general-purpose VLMs on medical VQA tasks in a zero-shot setting, without additional training.

Yim et al.~\cite{yim2024dermavqa} introduced DermaVQA, a dermatology-focused multilingual dataset, and benchmarked state-of-the-art VLMs on question-answering tasks without fine-tuning. Their findings highlight the models’ capability for cross-lingual response generation in dermatological contexts.

Ghosh et al.~\cite{ghosh2024clipsyntel} proposed a modular question summarization framework comprising four VLM-based components: medical disorder identification, relevant context generation, medical concept filtration, and visually-aware answer generation. Each module is tailored to a specific task within the pipeline, collectively improving response quality and clinical relevance.

In addition to clinical applications, recent studies have explored the potential of Medical VQA systems in educational and training contexts. Benitez et al.~\cite{benitez2024harnessing} examined the role of ChatGPT and similar LLMs in medical education, assessing their utility across various stages of medical training and instruction. Further, the performance of multiple VLMs, including the GPT-series, Claude-series, and Gemini-series, are evaluated on Japanese specialty board examinations in fields such as radiology, nuclear medicine, and interventional radiology~\cite{hirano2024gpt,oura2024diagnostic,watanabe2024role}. These studies highlight the emerging role of VLMs in supporting medical knowledge assessment and continuing professional development.

\textbf{Hybrid strategies}. Some endeavors adopt a hybrid strategy of input augmentation and task-specific fine-tuning for medical applications. 
For example, Hu et al. \cite{hu2024interpretable} fine-tune LLaMa2 for the medical domain by leveraging a large-scale, clinically driven dataset, constructed using GPT-4 extraction of key clinical information from radiology reports. 
It further enhances model reasoning by incorporating multi-relationship graph learning, integrating spatial, semantic, and implicit relationships to improve interpretability and accuracy in medical VQA.
Additionally, in real-world remote diagnosis scenarios, images provided by patients often contain excessive background elements and misaligned lesion areas, which can significantly degrade vision-language alignment during model training.
To address this, Li et al. \cite{li2024zalm3} propose ZALM3, which leverages an LLM to extract keywords from conversation history and employs GDINO (VLM) as a zero-shot object detector to identify the relevant objects or regions in patient images based on these keywords.
This approach is zero-shot and remains compatible with the fine-tuning process of various medical VLMs, ensuring adaptability across different models.

Liu et al. \cite{liu2023parameter} propose VQA-Adapter for CLIP to efficiently facilitate domain adaptation to the medical field and employ bilinear attention to integrate visual and textual features.
Chen et al. \cite{chen20243d} concatenate textual and visual features as input to Vicuna and employ LoRA to fine-tune it for generating the final output.
These two methods fall into feature extraction and task-specific fine-tuning.

Existing methods, relying on cross-modal pre-training and fine-tuning, struggle with accuracy due to data scarcity and limited integration of medical knowledge.
Liang et al. \cite{liang2024candidate} present Candidate-Heuristic In-Context Learning (CH-ICL) framework, which enhances VLMs with external knowledge to directly answer medical visual questions.
Their approach decomposes the VQA task into multiple steps, leveraging VLMs in various ways across input augmentation, feature extraction, supervision enhancement, and direct deployment.
First, they directly use Scispacy~\cite{neumann2019scispacy} and ChatGPT to extract medical-related named entities from pathology image captions, constructing a pathology terminology dictionary (input augmentation).
Second, they leverage the dictionary to fine-tune a cross-attention-based module, the Knowledge Scope Discriminator, to select several relevant knowledge scopes that replace the image's textual descriptions, assisting the VLMs in generating the final answer (feature extraction).
Third, they employ CLIP-based VLMs to extract features of image, question, and candidates answer list separately, which are used to calculate loss to supervise the selection of promising answers from the list of candidate options (supervision enhancement).
Finally, they integrate all generated information into a structured prompt and process it through frozen VLMs to generate the final answer (direct deployment).

\subsubsection{Discussion}
According to our research, VLMs has been widely employed in medical VQA task, due to their inherent ability to generate answers based on user prompts.
However, some challenges still limited their implementation in real-world scenarios.
(1) Class Imbalance in medical VQA datasets.
Although some studies have designed data augmentation strategies or leveraged VLMs to supplement samples for rare classes, these methods still present poor performance in unprecedented and uncommon diseases, making it difficult to develop a unified network that meets clinical requirements. 
Moreover, due to the presence of unmeaning synthetic samples generated by current augmentation techniques and the hallucination of VLMs, the reliability of answers produced by networks trained on these datasets remains doubtful.
(2) Limited reasoning ability on complex problems.
Several studies have reached the same conclusion: medical VQA models exhibit poor performance when handling exceptionally complex questions.
This may be due to the fact that, medical VQA often overlooks the importance of multi-modal fusion.
Effectively integrating question and image features can enhance model robustness and facilitate domain adaptation of VLMs, particularly when addressing ambiguous questions from patients and detailed, expert-level queries from clinicians.
(3) Low-quality generated answer in multi-round dialogue.
In complex open-ended VQA tasks, models often generate repetitive or off-topic answers in subsequent rounds due to the lack of explicit dependencies between questions in different rounds, and the absence of effective mechanisms to utilize key information from conversation history.
These challenges can be solved by introducing memory mechanisms, which allow models to retain historical context across multiple rounds, or further utilizing advanced VLMs with context-aware reasoning, long-term dependency modeling, which can further enhance the model’s ability to generate coherent and contextually relevant answers.

\subsection{Segmentation}
 
Medical image segmentation aims to accurately delineate meaningful anatomical or pathological regions, such as organs and lesions, within medical images to support disease diagnosis, treatment planning, and quantitative analysis. Formally, the task can be defined as a mapping function: $f:X \rightarrow Y$, where $X\in \mathbb{R}^{H \times W \times C }$ represents the input medical image, and $Y\in \{0,1,..., K\}^{H \times W }$ denotes the corresponding segmentation mask that assigns each pixel to one of $K$ semantic classes.

Despite recent advances, the performance of conventional image-only segmentation models is hindered by two key challenges. First, acquiring sufficient high-quality annotated datasets is difficult due to the labor-intensive and time-consuming nature of manual labeling by domain experts. Second, the inherent complexity and variability of medical images, particularly the presence of severe distortions, ambiguous boundaries, and intensity inhomogeneities in pathological regions, complicates accurate segmentation.

To address these limitations, recent studies have explored the integration of VLMs into medical image segmentation frameworks through cross-modal learning. VLMs can leverage textual guidance to generate pseudo-labels, enabling zero-shot and few-shot learning scenarios that significantly reduce the dependence on large-scale annotated datasets. Moreover, descriptive cues from clinical reports can guide lesion delineation by providing context-aware semantic priors.

The incorporation of VLMs into medical image segmentation offers multiple advantages: enhanced model interpretability, improved generalization in low-data regimes, and stronger alignment between imaging data and textual medical records. These benefits collectively contribute to the development of more robust and clinically applicable segmentation systems.

\subsubsection{Adapting VLMs to Medical Image Segmentation}
In medical image segmentation tasks, the adaptation strategies of VLMs encompasses feature extraction, supervision enhancement, task-specific fine-tuning, direct deployment and hybrid strategies..

\textbf{Feature Extraction}.
The attention mechanism is widely used to align visual and textual features for language-guided medical image segmentation. In this paradigm, visual features often serve as queries, while textual features are treated as keys and values within a cross-attention module. For instance, ~\cite{zhong2023ariadne, huemann2024contextual, li2024tp} apply this strategy to enable the image features to ``search" for the most relevant textual features, extracting matching semantic information that guides the segmentation task.

However, the inherent ambiguity of abstract medical descriptions, coupled with the presence of similar targets in images, can lead to mismatches when relying solely on single-direction cross-attention. To address these challenges, several studies ~\cite{zeng2024abp, hu2024lga, xie2024simtxtseg, li2024textmatch, bui2024visual} have proposed bilateral-attention mechanisms to effectively integrate both text and image information. Typically, the bilateral-attention module consists of two cross-attention modules, with visual and textual features alternately serving as the query, and the other as the key and value. The connection patterns of these two cross-attention modules are sequential connection \cite{zeng2024abp,li2024textmatch,bui2024visual} and parallel connection\cite{hu2024lga,xie2024simtxtseg}, respectively. For the sequential connection, the textual feature is initially enriched by visual information, which is then used to refine the visual features. In contrast, the parallel approach directly enhances the visual feature with the initial textual feature, without further refinement. 

Moreover, many medical image segmentation networks adopt multi-scale frameworks, progressively refining feature representations across different scales. The bilateral-attention mechanism can update textual features with multi-scale visual information at each layer, enhancing the optimization of high-level image features. Guo et al.~\cite{guo2024common} introduced an alternative bilateral-attention approach called common-attention, which treats both visual and textual inputs equally. Instead of assigning one modality as the query, this method projects both textual and visual features into keys and values through separate linear layers. The common attention matrix is computed using keys from both modalities, and this matrix is then used to generate a unified multi-modal feature. In conjunction with the Iterative Text Enhancement Module, which iteratively aligns textual features with image features across layers, this approach facilitates more effective deep inter-modal interaction in a multi-scale setting.

In terms of matrix operations, Han et al.~\cite{han2023multiscale} process word vectors through convolution blocks, which are subsequently added to image features, reconciling image and text information. In ~\cite{liu2024universal}, the global image feature and language embedding are concatenated and passed through a Multi-Layer Perceptron (MLP), which generates parameters corresponding to organ classes. This approach addresses the label orthogonality problem and improves performance in organ segmentation and tumor detection tasks.

Some studies integrate multiple operations to tackle more complex segmentation challenges. For example, Yue et al.~\cite{yue2023part} proposed SP-SAM, a text-promptable segmentation framework designed for surgical instrument segmentation. The Cross-Modal Prompt Encoder in SP-SAM aligns visual features with category-level and part-level textual collaborative prompts using structured attention and matrix transformations, generating part-aware embeddings for surgical instruments. By leveraging structured vision-language alignment, SP-SAM enhances segmentation accuracy, particularly in cases involving occlusions, fine-grained structures, and diverse instrument compositions.

Similarly, Zhou et al.~\cite{zhou2023text} also propose a text-promptable surgical instrument segmentation network. Their method first computes cross-attention to localize the instrument region in the visual features based on textual guidance. Then, a Text-Parameterized Convolution (TPC) operation is applied to transform the textual feature into convolutional layer parameters, which are used to convolve the enhanced visual features and generate score maps. Visual and textual features are concatenated and fed into a gating network to produce weights, which are used to compute the weighted sum of score maps to obtain the final segmentation results.

\textbf{Supervision Enhancement}.
Chen et al.~\cite{chen2024bi} proposed Bi-VLGM, a vision-language graph matching (VLGM) framework for text-guided medical image segmentation. The method addresses a critical limitation in classical VLMs, the distortion of intra-modal relationships (e.g., spatial relations between visual features or semantic consistency between textual features). VLGM introduces a structured supervision mechanism by constructing vision-language graphs and employing a graph matching strategy to ensure optimal cross-modal alignment. This design effectively preserves intra-modal structures while mitigating inconsistencies caused by direct cross-modal supervision. Additionally, Bi-VLGM incorporates both word-level and sentence-level alignment modules: the former aligns local image features with class-aware prompts, while the latter integrates global severity-aware prompts, thereby enriching the segmentation output with disease severity context.

\textbf{Task-specific Fine-tuning}.
Dhakal et al.~\cite{dhakal2024vlsm} introduced VLSM-Adapter for adapting large-scale Vision-Language Segmentation Models (VLSMs) to medical imaging tasks. Their method inserts multiple types of adapter modules at strategic points within CLIP-based transformer architectures to refine vision-language feature representations. On the other hand, Paranjape et al.~\cite{paranjape2024adaptivesam} proposed Bias-and-Norm-Tuning (BANT), a lightweight adaptation method for SAM. BANT incorporates a learnable bias parameter $\bar{b}$ into the output of each affine transformation layer within the transformer's architecture. Both studies demonstrate that adapter-based fine-tuning offers a scalable and computationally efficient solution for deploying VLMs in medical image segmentation, particularly in scenarios constrained by limited annotated data and computational resources.

\textbf{Hybrid Strategies}. 
Koleilat et al.~\cite{koleilat2024medclip} presented a hybrid segmentation strategy combining zero-shot inference and weak supervision. They leveraged BiomedCLIP~\cite{zhang2023biomedclip} and gScoreCAM~\cite{chen2022gscorecam} to extract bounding boxes based on input images and text prompts, which were then fed into SAM to generate initial pseudo-masks. These pseudo-masks were used as the supervision signals to train a separate segmentation network in a weakly supervised learning manner. This approach significantly improves the accuracy of zero-shot segmentation while maintaining minimal reliance on manual annotations.

\subsubsection{Discussion}
While VLMs have demonstrated promising capabilities in medical image segmentation, several challenges remain that limit their widespread clinical adoption. First, most existing approaches rely heavily on cross-attention-based fusion modules to integrate multi-modal visual and textual features. Although effective, these methods often incur substantial computational overhead, posing barriers to deployment in time-sensitive and resource-constrained clinical environments. As such, the development of efficient and lightweight fusion mechanisms tailored for real-world medical applications is a critical avenue for future research. Second, recent findings by Jiang et al.~\cite{jiang2024tv} reveal that the segmentation performance of VLMs on radiological datasets, particularly CT and MRI, is significantly lower than on non-radiological datasets. This performance gap may be attributed to the domain mismatch: non-radiological images, such as endoscopic or dermatological images, share more visual similarity with the natural image distributions on which most VLMs are pre-trained. Addressing this domain gap for radiological modalities remains a pressing challenge for enhancing VLM generalization in medical imaging. Lastly, while the incorporation of textual information has proven effective for guiding lesion localization and managing multi-scale complexity, it remains insufficient for resolving other critical segmentation challenges. In particular, VLMs still struggle with ambiguous boundaries, low-contrast regions, and other difficult visual scenarios. To enhance segmentation robustness, future work should explore complementary strategies, such as incorporating boundary-aware text prompts or integrating structural priors into the segmentation pipeline.

\subsection{Anomaly Detection}
 
Medical image anomaly detection focuses on identifying deviations from normal anatomical structures, such as tumors, lesions, fractures, or organ deformations within medical images. Formally, this task can be formulated as $f:X\rightarrow Y$, where $f$ denotes the deep learning model, $X$ represents the input medical image, and $Y$ is the corresponding anomaly label indicating the presence or absence of pathological anomaly. 
In unsupervised scenarios, the model additionally learns an anomaly score $S$, where higher values suggest a greater likelihood of abnormality.
Traditional deep learning methods face several challenges in this domain. The significant variability in the shape, texture, and size of pathological regions complicates model generalization. Furthermore, collecting large-scale, high-quality annotated datasets remains resource-intensive. Many existing approaches require training modality-specific models, leading to scalability issues as the number of imaging modalities and anatomical regions increases.

The integration of VLMs offers promising solutions to these limitations. By incorporating textual information and leveraging pre-trained knowledge, VLMs enable zero-shot or few-shot anomaly detection, reducing reliance on extensive annotations. Moreover, VLMs can unify anomaly detection across multiple imaging modalities within a single architecture, thereby enhancing cross-domain adaptability and simplifying deployment. Furthermore, incorporating VLMs into medical anomaly detection frameworks enables earlier identification of rare or underrepresented conditions, supports diagnostic generalization with limited data, and provides a scalable path forward for real-world clinical applications.

\subsubsection{Adapting VLMs to Anomaly Detection}
VLMs are primarily applied to feature extraction and task-specific fine-tuning in the context of medical anomaly detection, with relatively fewer studies exploring direct deployment without fine-tuning.

\textbf{Feature Extraction}.
Park et al.~\cite{park2024contrastive} proposed Contrastive Language Prompting (CLAP), a framework that employs both positive and negative textual prompts to guide attention-based anomaly detection. Positive prompts help enhance lesion localization, while negative prompts suppress attention to normal anatomical regions, thereby reducing false positives and improving detection precision. Lu et al.~\cite{lu2024multimodalf} introduced a stacked cross-attention module to effectively fuse image and text features extracted by a VLM, resulting in more comprehensive anomaly detection. To address scenarios where textual input may be unavailable, they also proposed an image-only branch. Performance knowledge is transferred from the multi-modal fusion model to this unimodal branch via knowledge distillation, ensuring robustness under varying input conditions.

In contrast, Chen et al.~\cite{chen2024clip} and Zhu et al.~\cite{zhu2024toward} employed matrix-based fusion techniques to combine visual and textual features. Their models, trained on industrial images and evaluated on medical imaging tasks, demonstrated strong zero-shot and few-shot generalization capabilities, underscoring the domain-transfer potential of VLM architectures even without extensive retraining on medical data.

\textbf{Supervision Enhancement}.
Sun et al.~\cite{du2024prompting} were the first to apply VLMs to dental abnormality detection, demonstrating accurate identification of multi-level anomalies in panoramic X-ray images. Their approach fine-tunes a VLM using multi-level prompting guided by the dental notation system. This strategy enables the detection of both global abnormalities across the entire oral image and local anomalies by aligning specific teeth, leveraging the inherent bilateral symmetry of the oral cavity for precise localization. MediCLIP~\cite{zhang2024mediclip} is a few-shot medical image anomaly detection framework that integrates multiple VLM utilization strategies. The method adapts CLIP through the use of learnable prompts and adapter-based fine-tuning, and computes cosine similarity between image and text embeddings to produce both anomaly scores and segmentation maps. In addition, Sun et al.~\cite{sun2024position}, as previously discussed, apply task-specific fine-tuning to adapt VLMs for anomaly detection, further demonstrating the flexibility of fine-tuning approaches in handling specialized medical imaging tasks.

\textbf{Direct Deployment}. Marzullo et al.~\cite{marzullo2024exploring} conducted a direct evaluation of CLIP's zero-shot generalization capability for detecting medical-specific anomalies, without applying any task-specific training. Their approach utilizes CLIP to project both medical images and text prompts that represent normal and abnormal conditions into a shared embedding space. Anomaly detection is then performed by computing the cosine similarity between image embeddings and the corresponding text-based descriptions. While CLIP exhibits a certain degree of knowledge transfer from its pre-training on natural image-text pairs, the model's performance remains below clinical standards, particularly in terms of diagnostic accuracy and sensitivity. These results underscore the necessity of domain-specific fine-tuning to adapt general-purpose VLMs for high-stakes medical applications.

\textbf{Hybrid Strategies}. 
Huang et al.~\cite{huang2024adapting} proposed an approach that not only fine-tunes the CLIP visual encoder for the medical domain using a multi-level visual feature adapter, but also integrates visual and textual embeddings through matrix product operations, enhancing cross-modal alignment for anomaly detection tasks. Similarly, Cao et al.~\cite{cao2025adaclip} adopted a combination of prompt tuning and matrix-based fusion, further demonstrating the effectiveness of these lightweight yet powerful techniques for improving medical anomaly detection performance in low-data or zero-shot settings.

\subsubsection{Discussion}
Despite recent advances, current VLM-based medical image anomaly detection methods still face several limitations. First, most existing approaches rely heavily on CLIP, which is primarily trained for global semantic alignment and often struggles to detect small or localized lesions. Although some studies have incorporated multi-scale feature extraction modules to alleviate this issue, there remains substantial room for improvement. Future research could explore the integration of alternative VLM architectures, potentially pre-trained with fine-grained medical supervision, to enhance spatial sensitivity and overall anomaly detection performance. Second, the textual input used in anomaly detection tasks is often highly homogenized. Many studies adopt minimalistic prompts such as ``normal/abnormal [object]", which fail to fully exploit the expressive capacity of VLMs. These limited textual formulations lack the rich semantic detail necessary for effective multi-modal fusion and precise localization. To address this, future work should consider enhancing textual descriptions with clinically relevant contextual cues, including lesion size, shape, texture, and anatomical location. Such fine-grained language can improve both the interpretability and accuracy of anomaly detection by guiding the model toward more targeted visual-semantic alignment.

\subsection{Medical Image Classification}
 
Medical image classification aims to categorize medical images into predefined diagnostic categories, supporting clinical decision-making processes such as disease diagnosis, prognosis assessment, and treatment planning. This task leverages computational models to extract and analyze discriminative features from medical imaging data. Mathematically, it can be formulated as a mapping function $f:X \rightarrow Y$, where $X $represents medical images, and $Y$ represents their corresponding labels.

Traditional deep learning-based approaches to medical image classification face several limitations: 1) High data dependency and annotation cost: These models typically require large-scale labeled datasets for effective training. However, annotating medical images demands domain-specific expertise, often involving board-certified radiologists, making the labeling process both time-consuming and expensive. 2) Limited interpretability: Deep learning models often operate as ``black boxes," providing limited insight into the rationale behind their predictions. This lack of transparency poses a significant barrier to clinical adoption, where explainability is essential for trust and validation. 

VLMs offer promising solutions to these challenges. First, they leverage self-supervised pre-training and vision-language alignment, allowing the model to learn semantically rich and generalizable feature representations from large volumes of unlabeled medical data, thereby reducing dependence on extensive manual annotation. Second, by integrating textual descriptions with visual features, VLMs enable concept-driven and interpretable classification, enhancing transparency and clinical relevance. The incorporation of VLMs into medical image classification frameworks has the potential to significantly improve diagnostic accuracy and enable automated, explainable clinical decision support, thus promoting more efficient and scalable healthcare delivery.

\subsubsection{Adapting VLMs to Medical Image Classification}

\textbf{Feature Extraction}.
VLMs have been widely used in medical image classification by extracting features from both images and text and integrating them through various fusion strategies to enhance diagnostic accuracy. 

A common approach is concatenation-based fusion, where extracted textual and visual features are combined directly. 
Monajatipoor et al. \cite{monajatipoor2022berthop} proposed BERTHop, which leverages PixelHop++ and VisualBERT to capture associations between clinical notes and chest X-ray images. 
Similarly, Huemann et al. \cite{huemann2023domain} explored domain-adapted BERT variants for nuclear medicine report classification, integrating them with vision models to improve PET/CT report analysis. 
Cao et al. \cite{cao2024mpoxvlm} introduced MpoxVLM, which combines CLIP’s visual encoder with LLaMA-based language models to analyze skin lesion images alongside patient data for mpox diagnosis.

Beyond simple concatenation, attention-based fusion enables deeper interactions between modalities \cite{song2022multimodal}. 
Meanwhile, a hybrid approach combining concatenation and attention mechanisms has been explored for classification tasks. 
Fatima et al. \cite{fatima2024transforming} designed a VQA-based model for microscopic blood cell classification, where textual embeddings and visual embeddings are fused using a dual-stream attention-concatenation strategy, demonstrating how VQA architectures can be repurposed for medical image classification.



\textbf{Supervision Enhancement}.
Several studies have leveraged VLMs to enhance supervision in medical image classification tasks either by refining ground truth annotations or by integrating VLM-derived supervision signals directly into model optimization through loss functions. An application of VLMs is to mitigate the impact of noisy or incomplete labels, which are common in large-scale medical datasets. Chen et al.~\cite{chen2023bomd} addressed this challenge in multi-label chest X-ray classification by optimizing a Bag of Multi-label Descriptors (BoMD). This technique aligns image predictions with VLM-generated semantic embeddings, thereby improving the consistency and reliability of noisy labels.

VLMs have also been employed for ground truth refinement and loss function enhancement. In some cases, VLMs are used to extract richer and more structured annotations, incorporating clinically relevant information such as anatomical location, severity levels, and diagnostic uncertainty. Other studies, such as Lanfredi et al.~\cite{lanfredi2025enhancing} and Lei et al.~\cite{lei2023clip}, have integrated textual knowledge derived from VLMs into contrastive learning objectives. This enhances the semantic granularity of the learned representations, improving classification performance and model generalization. The above-mentioned approaches demonstrate the utility of VLMs in strengthening both label supervision and optimization pipelines, ultimately enhancing the robustness, accuracy, and interpretability of medical image classification models.


\textbf{Task-specific Fine-tuning}.
Fine-tuning VLMs has emerged as a promising strategy to enhance diagnostic accuracy in medical image classification tasks by leveraging pre-trained multi-modal knowledge. Recent research has explored a range of fine-tuning methodologies aimed at improving model interpretability, robustness, and generalizability across diverse clinical applications. One notable direction involves ranking-aware learning, which improves performance in disease grading tasks by enforcing ordinal consistency in classification outputs. For instance, Yu et al.~\cite{yu2024clip} applied this approach to the assessment of diabetic retinopathy, achieving more accurate and semantically consistent severity predictions.

Beyond visual adaptation, fine-tuning strategies have also incorporated structured textual and numerical data to refine model understanding. Wang et al.~\cite{wang2024enhancing} demonstrated this in a video-based classification setting, where the integration of gait parameters and descriptive clinical text improved diagnostic reasoning and multi-modal alignment. Another critical advancement involves the use of domain-specific textual guidance. Zhang et al.~\cite{zhang2023text} embedded biomedical text representations during fine-tuning to enhance performance on histopathology datasets, particularly under data-scarce conditions. This approach reinforces domain alignment and supports better model generalization.

Fine-tuning has also been employed to improve model robustness in complex imaging scenarios. For example, Ding et al.~\cite{ding2024hia} demonstrated that adapting VLMs with high-resolution, instruction-aware modules enhances stability across variable input conditions. In a word, fine-tuning VLMs not only improves classification accuracy but also facilitates a deeper understanding of disease progression, allowing models to generate clinically meaningful insights that extend beyond categorical prediction~\cite{yu2024clip, ding2024hia}.

\textbf{Direct Deployment}.
The adoption of VLMs for medical image classification has gained momentum due to their ability to generate diagnostic output directly, without requiring task-specific training. This paradigm significantly reduces dependence on annotated datasets and enables zero-shot or in-context learning, facilitating broad applicability across diverse medical imaging domains.

Recent studies have evaluated state-of-the-art multi-modal VLMs, including GPT-4V, Gemini, and Claude 3 Opus, for direct classification in ophthalmology, dermatology, radiology, and oncology~\cite{upadhyaya2024360}. Ghalibafan et al.~\cite{ghalibafan2022applications} assessed GPT-4V for vitreoretinal disease classification, observing strong performance on structured prompts but limited generalization in open-ended cases.

In dermatological applications, Liu et al.~\cite{liu2024claude} benchmarked Claude 3 Opus and GPT-4V on melanoma detection, reporting moderate accuracy but notable difficulty in distinguishing between benign and malignant lesions. Ferber et al.~\cite{ferber2024context} demonstrated that in-context learning using GPT-4V can rival or even surpass task-specific deep learning models in cancer histopathology classification, showcasing the model’s zero-shot reasoning potential.

REMONI \cite{ho2024remoni} is an autonomous remote health monitoring system tailored to monitor chronic diseases and detect emergencies, including falls and anomalies in vital signs. Chen et al. \cite{chen2024applying} presents a smart telemedicine diagnosis system that integrates YOLOv7 \cite{wang2023yolov7} for object detection and ChatGPT to classify and diagnose pressure injuries, a common issue among bedridden and elderly patients.

Several studies have further benchmarked VLMs against radiologists and traditional deep learning models. Chen et al.~\cite{chen2024assessing} found that ChatGPT-4o and Claude 3 Opus underperformed compared to junior radiologists in thyroid nodule classification. Suh et al.~\cite{suh2024comparing} and Horiuchi et al.~\cite{horiuchi2024comparing} observed that GPT-4V and Gemini Pro Vision fell short in neuroradiology tasks, particularly in complex diagnostic cases. Than et al.~\cite{than2024comparison} compared several multi-modal LLMs with CNN-based models for tumor classification, concluding that although Gemini 1.5 Pro outperformed other VLMs, conventional deep learning models still yielded superior results when large annotated datasets were available.

To address the limitations of direct VLM deployment, recent efforts have explored prompt engineering as a means to enhance diagnostic accuracy. Guo et al.~\cite{guo2024performance} demonstrated that optimized prompt design can significantly boost VLM accuracy in tumor classification, although hallucination risks persist. Similar findings were reported by Yeh et al.~\cite{yeh2024insight} and Pellegrini et al.~\cite{pellegrini2023xplainer}, both of whom validated the effectiveness of prompt-based strategies for improving diagnostic precision in zero-shot medical classification.



\textbf{Hybrid Strategies}.
Recent research has demonstrated the effectiveness of combining multiple VLM strategies to enhance diagnostic performance in medical image classification. These hybrid approaches leverage the complementary strengths of VLM components across different categories to address key clinical challenges, such as data scarcity and feature misalignment.

Some methods integrate VLM-based generation and task-specific fine-tuning (Input Augmentation \& Task-specific Fine-tuning) to mitigate the limitations of limited annotated datasets. Chowdhury et al.~\cite{chowdhury2024adacbm} utilized GPT-4 to generate clinically relevant medical concepts for diagnostic networks and fine-tuned CLIP embeddings via an adapter module, enhancing both interpretability and classification accuracy. Other studies combine feature-level fusion (Feature Extraction) with supervision enhancement to improve the alignment between visual and textual features. Gao et al.~\cite{gao2024aligning} introduced a cross-attention mechanism to align diagnostic criteria embeddings from a VLM with image features, while also incorporating a semantic anchor loss to reinforce multi-modal consistency. Similarly, Yu et al.~\cite{yu2024language} and Zhu et al.~\cite{zhu2024stealing} applied this strategy in trauma classification and federated medical learning, respectively, further validating its generalizability across tasks. Song et al.~\cite{song2024pneumollm} proposed PneumoLLM, a VLM-based model for pneumoconiosis classification that integrates an MLP adapter and a novel classification head into a pre-trained backbone. By emphasizing classification outputs over complex textual generation, this framework simplifies the diagnostic pipeline and enhances adaptability in data-scarce environments (Feature Extraction \& Task-specific Fine-tuning).

\subsubsection{Discussion}
Despite the increasing adoption of VLMs in medical image classification, their clinical applicability remains limited by several critical challenges. A foremost issue lies in their vulnerability to performance degradation on underrepresented classes, especially within long-tailed distributions, a common characteristic in specialized medical datasets. This limitation is further exacerbated by pervasive data scarcity in rare disease domains. While traditional techniques such as static data augmentation or loss reweighting have been employed to alleviate class imbalance, recent advances highlight the importance of dynamic training strategies, including curriculum-guided sampling, which progressively shifts learning focus from majority to minority classes. In addition to these strategies, developments in generative modeling now enable targeted image synthesis of pathologically relevant samples. Additionally, text-guided image editing techniques allow fine-grained modification of medical images based on diagnostic descriptions, further enriching training distributions with clinically meaningful variations. However, these data-centric solutions alone are insufficient to address challenges arising from domain shifts during cross-modal knowledge transfer. To improve model robustness, architectural innovations are necessary, particularly those that disentangle domain-invariant representations from modality-specific perturbations. Integrating uncertainty-aware mechanisms, such as dynamic decision boundaries based on feature-space density estimation, can further enhance adaptability in real-world diagnostic settings. Moving forward, future research should focus on the co-design of synthetic data pipelines and adaptive model architectures. In particular, enforcing cross-modal consistency constraints between visual and textual embeddings holds promise for enhancing alignment, improving generalization, and ultimately advancing the clinical viability of VLMs in medical image classification.

\subsection{Medical Image Synthesis}
 
Medical image synthesis aims to generate images of a target modality from a given source modality to address the challenges associated with acquiring multiple imaging modalities in clinical workflows. This task plays a critical role in enhancing diagnostic completeness and treatment planning, especially in scenarios where access to certain imaging modalities, such as PET or high-resolution MRI, is limited due to high acquisition costs or patient safety concerns. However, medical image synthesis remains inherently challenging due to the complex and non-linear mappings between different imaging domains. Traditional approaches often rely on handcrafted features, which limit their capacity to model fine-grained anatomical detail and to generalize across diverse patient populations and imaging conditions. In contrast, modern deep learning-based approaches, notably Generative Adversarial Networks (GANs)~\cite{goodfellow2014generative} and Denoising Diffusion Probabilistic Models (DDPMs)~\cite{ho2020denoising}, have demonstrated the ability to learn complex inter-modal relationships and produce high-fidelity synthetic images.
Formally, given input $x \in \{ I,T,(I,T)\}$, where $I$ represents the source image modality and $T$ represents textual or descriptive information, the image synthesis network aims to generate an image $y$ in the target modality. By enabling the generation of clinically relevant and highly detailed images, such models have the potential to reduce radiation exposure, lower healthcare costs, and expand access to advanced imaging, for example, in MRI-only radiotherapy planning or PET/MRI hybrid scanning.

\subsubsection{Adapting VLMs to Medical Image Synthesis}
The application of VLMs in medical image synthesis primarily concentrates on Task-specific Fine-tuning, followed by Feature Extraction and Input Augmentation. Several studies also integrate multiple strategies to form a hybrid solution. Detailed implementations are as follows:

\textbf{Input Augmentation}. Chen et al.~\cite{chen2024medical} not only leverage GPT-4 to construct detailed anatomy-pathology descriptive prompts, but also integrate multiple frozen VLMs to extract textual and image features, finally extracting key patches as visual cues to generate high-quality medical images.

\textbf{Feature Extraction}. Xu et al.~\cite{xu2024medsyn} and Hamamci et al.~\cite{hamamci2025generatect} utilize cross-attention mechanisms within DDPM-based modular frameworks to integrate textual and image features. Xu et al.~\cite{xu2024medsyn} synthesize anatomically precise 3D lung CT images by combining textual prompts with segmentation masks, effectively addressing memory constraints while ensuring anatomical plausibility for downstream applications. Building on this, Hamamci et al.~\cite{hamamci2025generatect} rely solely on freeform text prompts to generate high-resolution 3D medical images, demonstrating its effectiveness as a data augmentation technique for multi-abnormality classification tasks. Additionally, Lee et al.~\cite{lee2023vision} concatenate multi-view chest X-ray embeddings with report embeddings as the input sequence to generate specific views of chest X-rays. This integration helps rectify potential errors and enhances the ability to faithfully capture abnormal findings in chest X-ray generation. Notably, ~\cite{bahani2024enhancing} introduces an enhanced DF-GAN architecture that leverages affine transformations to deeply integrate textual descriptions and image features for generating synthetic chest X-ray images. The employment of affine transformations enables precise control over how text features influence image generation, ensuring semantic consistency and improving the fidelity of the synthesized images.

\textbf{Task-specific Fine-tuning}. Some DDPM-based networks inherently generate natural images from text prompts, and researchers have further fine-tuned them for medical applications. HistoSyn et al.~\cite{yin2024histosyn} leverage histomorphological-focused text prompt to fine-tune a Latent Diffusion Model (LDM)~\cite{rombach2022high} to synthesize diverse and realistic pathology images. 
Unlike conventional text prompts, HistoSyn generates its prompts by extracting spatial and morphological features from key objects (e.g., nuclei, fat droplets) segmented from pathology images, effectively capturing fine-grained histological details tailored to the task. This design ensures that the generated images preserve diagnostic attributes, while improving performance in downstream pathology-related tasks.
Shavlokhova et al. \cite{shavlokhova2023finetuning} fine-tune the GLIDE \cite{nichol2021glide} using 10,015 dermoscopic images annotated with seven diagnostic entities to generate high-quality synthetic images tailored for dermatological applications.
The fine-tuning for specific datasets easily and effectively  improves the  performance of generating particular modality images to address data scarcity.

\textbf{Feature Extraction \& Task-specific Fine-tuning}. The challenges associated with medical image synthesis have led many researchers to adopt and integrate multiple strategies to enhance the quality and applicability of generated images. The integration of cross-attention mechanisms to combine CLIP-generated text embeddings with image features in the Denoising U-Net, alongside the direct fine-tuning of Stable Diffusion as described in \cite{bluethgen2024vision}, has proven effective for generating high-quality CXR images. This combined approach achieves a balance between semantic precision from text-guided synthesis and enhanced realism through fine-tuned model adaptation, resulting in superior image alignment and fidelity.

\subsubsection{Discussion}
Medical image synthesis represents a rapidly evolving research area with the potential to deliver substantial benefits across diverse clinical applications. By leveraging the prior knowledge embedded in VLMs, numerous studies have demonstrated improved alignment between textual prompts and visual outputs, enabling the generation of high-quality, semantically relevant images guided by user-defined descriptions. 

Despite this progress, several critical challenges persist in applying VLMs to medical image synthesis: First, current research predominantly focuses on X-ray image generation, with relatively limited attention given to other widely used clinical modalities, such as MRI and PET. This limitation is primarily due to the scarcity of open-source datasets that pair these imaging modalities with corresponding textual annotations. However, with the ongoing advancement of medical-specific VLMs, automatically generating rich, structured text descriptions for underrepresented modalities presents a promising solution to bridge this data gap. Second, the field exhibits a heavy reliance on Denoising Diffusion Probabilistic Models (DDPMs) for image generation. While DDPMs have shown exceptional capability in producing diverse and detailed synthetic images, they suffer from notable limitations, including long inference times and a susceptibility to generating artifacts or ``hallucinations" that compromise image realism. These drawbacks hinder the scalability, efficiency, and reliability of VLM-assisted synthesis, particularly in time-sensitive or precision-critical clinical contexts. Third, there exists a fundamental modality gap between the information density of medical images and the expressive capacity of textual descriptions. Medical images inherently contain high-resolution spatial and morphological details that are often difficult to fully capture in natural language. This semantic disparity can result in incomplete, ambiguous, or anatomically inconsistent synthetic images, thereby limiting their diagnostic utility and clinical relevance. 

In summary, while VLM-based medical image synthesis has shown considerable promise, the field remains in its early stages. Addressing the current limitations, through dataset diversification, architectural optimization, and enhanced cross-modal representations, will be essential for advancing this technology toward reliable, real-world clinical deployment. Moreover, unexplored modalities and hybrid synthesis paradigms offer fertile ground for future investigation.

\subsection{Medical Image-Text Retrieval}

The primary objective of medical retrieval is to retrieve clinically relevant images or textual information based on a given query, such as an image, report, or natural language description, to support diagnostic decision-making, enhance clinical workflow efficiency, and facilitate medical research and education.
This task can be formulated as $\hat{I} = \arg\max_{I \in D} S(Q, I)$ where $Q$  represents the query (image or text), $ D $ is the database,  $I$  is a candidate item in the database, and  $S(Q, I)$ is the similarity function that ranks the retrieved items based on their relevance.
Traditional deep learning-based medical retrieval methods face two critical limitations. First, they exhibit limited multi-modal understanding, often struggling to effectively integrate image and textual data. This deficiency reduces the system’s capacity to deliver context-rich and semantically aligned retrieval results. Second, such models frequently encounter generalization challenges, where performance degrades significantly when applied to unseen medical domains or datasets due to overfitting to narrow training distributions.

The integration of VLMs offers promising solutions to these challenges. By enabling joint representation learning across visual and textual modalities, VLMs facilitate context-aware retrieval whether the query is an image, clinical note, or free-form description. Additionally, their foundation in large-scale pretraining improves generalizability across varied medical tasks and imaging modalities, even in low-resource or domain-shift scenarios. The application of VLMs to medical retrieval yields tangible benefits, including enhanced diagnostic decision support, more accurate cross-modal data matching, and improved efficiency in clinical information management. Collectively, these capabilities contribute to better patient outcomes, accelerated medical research, and a more intelligent, integrated healthcare infrastructure.

\subsubsection{Adapting VLMs to Medical Image-Text Retrieval}
\textbf{Feature Extraction}. The attention mechanism, renowned for its adaptability across a range of machine learning tasks, has been effectively applied to medical image and text retrieval. In particular, multi-head attention (MHA) plays a critical role in refining cross-modal feature fusion and enhancing retrieval relevance. For example, X-TRA~\cite{Van2023x} integrates retrieved representations from CLIP and PubMedCLIP using a FAISS-based retrieval index. These retrieved features, representing semantically similar prior cases, are incorporated via an MHA module to enhance disease classification and clinical report retrieval. The attention mechanism dynamically weights the importance of each retrieved representation, allowing the model to emphasize diagnostically relevant features while filtering out noise. By leveraging CLIP-encoded prior knowledge through attention-guided fusion, this approach significantly improves retrieval accuracy and interpretability, enabling more context-aware decision support in medical applications.

\textbf{Task-specific Fine-tuning}.
Loss-based fine-tuning serves as a critical strategy for improving medical image-text retrieval by explicitly aligning multi-modal representations within a shared embedding space through task-specific objective functions. Shukor et al.~\cite{shukor2024vision} proposed a dual-loss framework that combines Image-Text Contrastive (ITC) loss, which employs triplet learning to separate mismatched image-text pairs while clustering aligned ones with Image-Text Matching (ITM) loss, which performs binary verification to assess the semantic correctness of each pair. This dual-loss strategy facilitates both coarse-grained and fine-grained alignment, enhancing retrieval robustness. Windsor et al.~\cite{windsor2023vision} introduced a hierarchical loss framework, integrating multiple alignment levels. Their model employs InfoNCE~\cite{oord2018representation} for global alignment, GLoRIA~\cite{huang2021gloria} for local alignment between text segments and image regions, and utilizes DeCLIP-generated augmented and nearest-neighbor pairs to compensate for data scarcity. This multi-level supervision reinforces the model’s ability to handle diverse medical scenarios. Despite methodological differences, both approaches demonstrate that tailored loss functions effectively guide models to map medical images and texts into a semantically consistent embedding space, where relevant image-text pairs cluster closely. This results in significantly enhanced retrieval performance across varied real-world clinical settings.

\textbf{Hybrid Strategies}.
When multiple utilization strategies of VLMs are combined for medical image-text retrieval, the most frequently employed is Input Augmentation. For example, GPT-4 is utilized in Chen et al.~\cite{chen2024bimcv} to translate Spanish radiology reports into English, enhancing cross-lingual accessibility and consistency across datasets. Similarly, CheXagent~\cite{liu2024medical, abacha20233d} generates additional textual captions for medical images, enriching semantic context and improving image-text alignment. Apart from Input Augmentation, several studies incorporate other categories to strengthen different components of the retrieval pipeline. During the multi-modal fusion stage, MedFinder~\cite{chen2024bimcv} adopts the Task-Specific Fine-tuning strategy, implementing a cross-modal matching strategy, comparing image and text embeddings via cosine similarity to enable more effective text-to-image and image-to-text retrieval. Similarly, Liu et al.~\cite{liu2024medical} extract features from auto-generated captions using an adapter network to create contextual prompts. These are then used to fine-tune a VLM, aligning embeddings for both image and text queries. A noise-robust contrastive learning strategy is also employed to further improve retrieval accuracy. Abacha et al.~\cite{abacha20233d} employ the Direct Deployment strategy, extending retrieval capabilities to volumetric imaging by encoding captions of 3D medical images in tandem with 2D slice representations using BiomedCLIP. This approach creates a shared embedding space across text and 3D image inputs.
These above-mentioned hybrid strategies demonstrate the flexibility and effectiveness of VLMs in advancing medical image-text retrieval, offering improvements in accessibility, contextual understanding, and cross-modal alignment.

\subsubsection{Discussion}
VLMs have demonstrated significant promise in medical image-text retrieval by enabling robust multi-modal alignment and enhancing the interpretability of diagnostic queries. However, several critical limitations hinder their clinical integration and scalability in real-world healthcare environments: 1) High Computational Demands: The processing of high-resolution medical images using VLMs imposes considerable computational overhead, which restricts their applicability in real-time or resource-constrained clinical settings~\cite{garcia2024neui}. To address this, future efforts should explore lightweight model architectures, including model compression, adapter-based pruning, and cascaded processing pipelines that emphasize region-of-interest (ROI) prioritization, thereby minimizing redundant operations without sacrificing diagnostic performance. 2) Insufficient Domain-Sensitive Representation: Many VLMs struggle to capture fine-grained, domain-specific medical features due to limitations in cross-modal alignment and reliance on global feature extraction. This often results in the omission of clinically critical regions, such as small skin lesions or localized organ anomalies. Enhancing local-region feature fusion and incorporating expert-guided annotation signals, either during training or prompt formulation, can improve sensitivity to subtle diagnostic indicators and enrich the semantic granularity of retrieved content. 3) Ineffective Long-Context Processing: Sequential or longitudinal medical data (e.g., temporal imaging series or evolving clinical notes) pose challenges for VLMs in terms of maintaining focus and coherence across extended multi-modal inputs~\cite{sharma2024losing}. Solutions such as modular attention mechanisms, hierarchical encoding schemes, and linear attention variants can optimize contextual retention while reducing distraction from irrelevant or redundant inputs. In summary, while VLMs hold substantial potential for advancing medical image-text retrieval, future research should focus on adaptive architectures that jointly optimize computational efficiency, granular medical feature extraction, and context-aware reasoning, thereby enabling seamless integration into end-to-end clinical workflows.

\subsection{Medical Multi-task Learning}
 
Medical Multi-Task Learning (MMTL) aims to develop a unified deep learning framework capable of simultaneously addressing multiple downstream medical tasks, such as classification, segmentation, and report generation, to enhance efficiency, diagnostic accuracy, and cross-task generalizability in clinical applications.
Mathematically, given an input $X \in \{I, T, (I,T)\}$, MMTL optimizes multiple tasks 
$T_i$ using a model $F$ and outputs the corresponding results $\{Y_i\}_{i=1}^{m} = F\{T_i\}_{i=1}^{m}(X)$, where the total loss function is formulated as $L_{\text{MMTL}} = \sum_{i=1}^{m} w_i L_i$, where $I, T$ represent the image and text inputs and $w_i$ denotes the loss weight of the $i$-th tasks.

Conventional deep learning-based approaches to MMTL often treat each task in isolation, failing to explicitly model inter-task correlations. This limits the potential to exploit shared semantic or anatomical features, which could otherwise enhance learning through synergistic task relationships. Furthermore, optimizing multiple tasks concurrently presents inherent challenges. For example, imbalanced loss weighting may cause certain tasks to dominate the training process, while others receive insufficient attention, ultimately degrading overall system performance. The integration of VLMs offers a transformative solution to these challenges. By leveraging shared latent spaces derived from multi-modal medical data (e.g., radiological images and clinical reports), VLMs can bridge the semantic gap across tasks, promoting better task correlation modeling. Moreover, VLMs support adaptive loss balancing, dynamically adjusting task-specific weights based on learning signals or task importance, thereby improving multi-task optimization efficiency~\cite{guo2024boosting}. These capabilities position VLMs as a promising foundation for building unified diagnostic systems that incorporate diverse medical tasks, such as detection, classification, segmentation, and report generation, within a single framework. This advancement not only contributes to treatment planning but also enables scalable, real-world deployment of intelligent clinical decision-support systems.

\subsubsection{Adapting VLMs to Medical Multi-tasks Learning}

\textbf{Feature Extraction}.
Recent studies in MMTL have increasingly adopted VLMs using feature extraction and fusion paradigms, wherein customized modules integrate textual and visual representations to support a range of downstream medical tasks. For example, Bai et al.~\cite{bai2023cat} introduced a Guided-Attention Module to fuse image and text features, enhancing VQA accuracy and spatial localization in surgical environments by enabling more effective multi-modal interaction. Similarly, Zhao et al.~\cite{zhao2024cap2seg} generated medical image captions as intermediate textual features, which interact with visual representations through a scale-aware attention mechanism, facilitating high-quality segmentation without requiring textual inputs during inference. In the domain of brain imaging, Wang et al.~\cite{wang2024ftspl} combined fMRI signals, textual inputs, and anatomical region embeddings via Pearson correlation, followed by graph neural network (GNN) processing. This fusion strategy extended VLM applicability to disease classification and cognitive prediction in neuroimaging. For cerebral microbleed detection, Chen et al.~\cite{chen2024novel} employed a concatenation-based fusion approach, directly integrating visual and textual features to enhance multi-modal representation learning, leading to improved detection and classification accuracy. In the context of chest radiography, Müller et al.~\cite{muller2025chex} utilized VLMs in two complementary ways: (1) CLIP-derived image-text embeddings were integrated into OV-DERT~\cite{zang2022open}, enabling multi-prompt-based bounding box prediction, and (2) multi-modal embeddings were applied to a GPT-2-based generator to produce textual descriptions for each region of interest (ROI). This dual approach enhances localized interpretability by associating textual prompts with specific visual regions, while also supporting region-level report generation, fostering more interactive and explainable diagnostic workflows. In summary, although these studies implement a variety of fusion strategies, including attention mechanisms, correlation-based integration, and concatenation, they uniformly exploit VLMs for multi-modal representation learning. These approaches demonstrate the potential of VLMs to advance medical MMTL frameworks by effectively bridging image-text modalities in complex clinical tasks.

\textbf{Supervision Enhancement}. Guo et al.~\cite{guo2024boosting} integrated VLMs into a Weakly Semi-Supervised Learning (WSSL) framework to address the dual challenges of tumor segmentation and cancer detection. The proposed method incorporates dynamic loss balancing, enabling the model to adaptively adjust the optimization focus between the two tasks, thereby improving training stability and convergence. By utilizing clinical reports as a source of weak supervision, the VLM component enhances diagnostic precision by mitigating errors caused by sparse or noisy annotations. Specifically, the model leverages text-derived contextual cues to reinforce critical spatial information related to tumor localization. This capability significantly reduces the reliance on expert-generated pixel-level annotations, while maintaining high diagnostic performance. Overall, this approach demonstrates a cost-effective and scalable solution for AI-driven cancer screening, offering strong potential for clinically viable deployment in resource-constrained environments.

\textbf{Task-specific Fine-tuning}. 
Several recent studies have explored comprehensive fine-tuning strategies to fully harness the potential of VLMs for unified MMTL, aiming to align these models more closely with the practical demands of clinical workflows. Nath et al.~\cite{nath2024vila} introduced Instruction Fine-Tuning (IFT) guided by expert model feedback, enhancing VLM performance across a variety of downstream tasks, including segmentation, classification, VQA, and medical report generation. This adaptive approach enables the model to dynamically access and apply specialized medical knowledge on demand, improving its generalizability and clinical relevance. Similarly, MiniGPT-Med~\cite{alkhaldi2024minigpt} utilizes LoRA~\cite{hu2021lora} and fine-tuning of the linear projection layer within the LLaMA-2 architecture~\cite{touvron2023llama} to support radiology-specific tasks such as disease identification, medical VQA, and report generation. This lightweight fine-tuning strategy facilitates efficient adaptation to high-resolution medical imaging and improves multi-modal integration with minimal computational overhead. Additionally, Bannur et al.~\cite{bannur2024maira} adopted similar fine-tuning-based approaches in developing MAIRA, a VLM-driven multi-task framework tailored for chest X-ray interpretation, demonstrating improved performance across multiple diagnostic subtasks. These studies illustrate that fine-tuning VLMs for MMTL offers a promising path toward building clinically robust, task-flexible AI systems, capable of delivering comprehensive diagnostic insights in real-world medical environments.

\textbf{Direct Deployment}.
Xu et al.~\cite{xu2024sat} introduced SAT-Morph, a framework that directly leverages the Segment Anything with Text (SAT) model~\cite{zhao2023one} to segment anatomical structures based on text prompts. These segmentation results are used to generate pseudo mask labels, which subsequently guide the training of a medical image registration model. By explicitly incorporating anatomical knowledge into the alignment process, SAT-Morph minimizes deformation inconsistencies and delivers a more robust and anatomically aware solution for image alignment tasks.

In another line of work, Manjunath et al.~\cite{manjunath2024towards} employed GPT-4 to generate diagnostic reports and facilitate question answering (Q\&A) for brain tumor classification within an image retrieval system. While the integration of a VLM enhances interpretability through summarization and interactive querying, the system lacks a fully integrated framework connecting retrieval, classification, and report generation. This disjointed design limits its ability to fully exploit multi-modal information synergy for comprehensive diagnostic support.

\subsubsection{Discussion}
VLMs have demonstrated substantial potential in MMTL by enhancing interpretability and supporting a broad range of diagnostic tasks. Nevertheless, several key limitations persist in applying VLMs effectively within multi-task clinical workflows. First, VLMs are frequently relegated to the role of a supplementary text-generation module, contributing only to report generation or language-based outputs, while remaining detached from core tasks such as segmentation or classification. This limited integration restricts their utility in spatially grounded tasks. To address this, the adoption of feature fusion mechanisms is essential, allowing VLMs to actively participate in spatial reasoning and visual representation learning, thereby enriching their role beyond narrative description. Second, current VLMs often struggle with complex instruction parsing, particularly in scenarios requiring multi-step reasoning, such as: ``Locate the tumor, then assess its stage.” As noted by Ying et al.~\cite{ying2024mmt}, this deficiency hampers VLMs' ability to follow structured medical protocols. Potential solutions include the use of hierarchical prompting strategies or the deployment of multi-agent collaborative architectures, wherein subtasks are decomposed and sequentially executed to preserve logical consistency and task relevance. Future research should prioritize enhancing VLMs’ capabilities in modeling task dependencies, coordinating inter-task interactions, and dynamically adapting to task context. These improvements are critical to evolving VLMs from modular components into truly unified, intelligent systems capable of supporting end-to-end medical decision-making across complex clinical pipelines.

\subsection{Other Image Analysis Tasks}

\subsubsection{Medical Image Super-resolution} 

Medical image super-resolution (SR) aims to enhance the spatial resolution of medical images, thereby improving the visualization of anatomical structures and disease-specific features. Formally, given a low-resolution (LR) input $X_{LR}$, the objective of SR is to produce a super-resolved output $Y_{SR}$ such that $Y_{SR} \approx X_{HR}$, where $X_{HR}$ represents the corresponding high-resolution (HR) image. In clinical practice, the quality of medical images is often limited by hardware constraints, resulting in low-resolution outputs that obscure fine-grained details essential for accurate diagnosis and treatment planning.

Traditional approaches, such as interpolation-based methods, are inadequate for restoring high-frequency information, which is critical for preserving diagnostic relevance. In contrast, deep learning-based SR models can infer and reconstruct missing high-frequency components, yielding sharper and more detailed reconstructions~\cite{lin2024dual}. These methods enhance the visibility of subtle anatomical features, supporting clinicians in early abnormality detection and improving diagnostic accuracy. However, conventional deep learning SR approaches are typically limited to image-only supervision and pixel-wise loss functions, which restrict their ability to capture complex texture patterns and semantic structures inherent in medical data. To overcome these limitations, integrating textual information through VLMs presents a promising avenue. VLMs offer semantic guidance that can enhance both perceptual quality and optimization stability by providing complementary contextual knowledge, thereby mitigating issues related to detail loss and over-smoothing.

Despite this potential, the application of VLMs in medical image SR remains underexplored. Notably, Ni et al.~\cite{ni2024m2trans} introduced M2Trans, a framework designed for ultrasound SR that leverages joint image-text supervision. Specifically, the model adopts the Supervision Enhancement strategy, utilizing the image and text encoders from MedCLIP~\cite{wang2022medclip} to compute an image-text semantic loss based on cosine similarity. In addition, M2Trans integrates the Caption Anything model~\cite{wang2023caption} to extract attribute-level textual features, which are embedded into templated sentences to form descriptive textual inputs for guiding the SR process, an approach falling into the Input Augmentation strategy. This hybrid supervision framework enhances semantic fidelity and improves texture reconstruction, revealing the untapped potential of VLMs in advancing the field of medical image super-resolution.

\subsubsection{Medical Image Registration} 

Medical image registration aims to estimate an optimal spatial transformation that aligns anatomical structures between a fixed image $X_{F}$ and a moving image $X_{M}$. This task is essential for accurately aligning scans from different modalities, correcting for anatomical variability, and enabling the interpretation of functional or structural changes across time or imaging protocols. In clinical settings, precise registration is foundational for tumor tracking, surgical planning, and monitoring disease progression or treatment response.

Traditional image registration approaches typically rely on iterative, non-convex optimization, which is computationally expensive and prone to convergence toward suboptimal local minima. To address these limitations, deep learning-based registration methods are proposed to enable the prediction of deformation fields $\phi$ through a single forward pass of a trained neural network. This deformation field is then applied to the moving image via spatial transformation, yielding an aligned output that closely approximates the fixed image.

Despite the success of deep learning in accelerating registration, the integration of VLMs into this domain remains relatively unexplored. A notable exception is the work of Chen et al.~\cite{chen2024spatially}, who proposed textSCF, a framework that leverages textual anatomical prompts encoded via CLIP to guide the registration process. This approach falls into the Feature Extraction strategy.

In textSCF, text embeddings are used to encode anatomical labels (e.g., segmentation masks), which are subsequently transformed into spatially covariant filter weights. These weights modulate the registration network’s feature maps, allowing the system to dynamically preserve structural discontinuities and enhance anatomical correspondence, particularly in challenging cases involving large deformations or limited training data. The incorporation of text not only introduces external anatomical priors but also improves the model’s generalizability and semantic awareness in structure-guided registration tasks.

\subsubsection{Object Detection}

Medical image object detection focuses on identifying and localizing clinically relevant objects or abnormalities, such as lesions, tumors, or anatomical landmarks within medical images to support diagnosis, treatment planning, and interventional guidance. This task presents significant challenges due to the inherent complexity, heterogeneity, and low inter-class variability of medical imaging data, along with the demand for high precision and real-time inference in clinical environments. Deep learning-based detection methods have demonstrated substantial advantages by automating feature extraction, improving detection accuracy, and enabling end-to-end training pipelines. Moreover, their capacity for real-time processing makes them particularly suitable for integration into point-of-care diagnostic systems and surgical navigation platforms.
Given an image $X$, the medical image object detection process can be formulated as $f:X\rightarrow Y$, predicting target object $Y=\{ (b_i, c_i, s_i )\}^N_{i+1}$ , where $b_i=(x_i, y_i, w_i, h_i)$ represents the bounding box coordinates, $c_i$ denotes the class label, $N$ is the number of detected objects, and $s_i$ indicates the confidence score for each detected object.
The clinical significance of medical image object detection lies in its ability to improve diagnostic accuracy, reduce human error, and enhance patient outcomes through the early and precise identification of disease-specific abnormalities. In contemporary AI-assisted diagnostic pipelines, object detection frequently serves as a foundational module within multi-task learning frameworks, supporting downstream tasks such as segmentation, classification, and report generation (refer to Section 4.8 for details).

Despite its central role, the use of VLMs for standalone object detection remains underexplored. Most existing research emphasizes zero-shot learning, leveraging the cross-modal generalization capabilities of VLMs to reduce dependence on large-scale, expert-annotated datasets.

Wu et al.~\cite{wu2023zero} tackled the task of nuclei detection using a multi-round prompt refinement strategy. Their method iteratively adjusts text prompts based on feedback from prior detection results, aligning prompt semantics more closely with underlying visual features. By incorporating fine-grained anatomical and morphological details, the approach reduces dependency on manually designed prompts and markedly improves visual-language alignment and zero-shot detection performance. This strategy exemplifies the Input Augmentation paradigm, as it enhances model guidance through dynamic textual input refinement.

Similarly, Guo et al.~\cite{guo2023multiple} explored zero-shot lesion detection without the need for extensive expert annotations. Their work represents the Direct Deployment strategy, leveraging pretrained VLMs in inference without additional task-specific training. To overcome domain adaptation challenges, particularly the difficulty of capturing medical-specific features like lesion shape, color, and size, the authors proposed a multi-prompt design, with each prompt focusing on a distinct lesion characteristic. The model outputs were then aggregated via an ensemble learning framework that integrated step-wise clustering and majority voting, yielding more robust and precise detection results across diverse clinical cases.

\subsubsection{Discussion}
Overall, the application of VLMs to downstream medical imaging tasks remains in its early stages, making it premature to draw definitive conclusions regarding their full potential and limitations. While object detection is commonly integrated as a sub-component within multi-task learning frameworks, the exploration of standalone medical object detection tasks remains relatively limited. Nevertheless, the downstream architectures reviewed in this work highlight the promising role of VLMs and text-based supervision in addressing critical challenges across a variety of medical tasks. These initial findings establish a strong foundation for future research and model refinement. Given the rapid advancements in VLM-assisted methodologies within the broader field of computer vision, a similar trajectory of innovation and adoption is anticipated in the medical domain in the near future.

\section{Challenges}
\label{sec:challenges}

While vision-language models (VLMs) hold significant promise for medical image analysis, their effective adaptation to downstream clinical tasks remains constrained by several technical and practical barriers. 
These challenges span data limitations, cross-modal integration issues, domain transfer gaps, and deployment feasibility in real-world settings.

\subsection{Data Scarcity}
The development of medical VLMs is fundamentally hindered by the limited availability of large-scale, high-quality image-text paired datasets~\cite{khattak2024unimed}.
Acquiring medical images, especially high-resolution 3D scans or multi-sequence modalities such as MRI and CT, entails considerable cost, specialized equipment, and protocol heterogeneity across institutions~\cite{bahloul2024advancements}. 
Furthermore, annotation remains a substantial bottleneck: pixel-level segmentation or structured diagnostic labeling requires expert radiologists and pathologists, making scalable dataset curation labor-intensive and time-consuming~\cite{lai2025med}.

On the other hand, textual modalities suffer similar constraints. 
Many public datasets either lack associated clinical narratives or rely on oversimplified descriptions, such as generic template-based captions~\cite{li2024vclipseg,sun2024position}. 
Even when radiology reports are available, they often contain dense medical jargon, abbreviations, and implicit reasoning, posing challenges for accurate semantic parsing and alignment. 
Moreover, different downstream tasks impose divergent requirements on textual inputs~\cite{lai2025med}. 
For example, image segmentation demands spatially grounded descriptions, while disease classification emphasizes morphological and contextual cues~\cite{gao2024aligning}. 
These limitations significantly impact Input Augmentation and Supervision Enhancement, which rely on high-quality data to synthesize informative inputs or training signals. Poor-quality or sparse source data can lead to unreliable augmentation and supervision, introducing noise and reducing downstream model performance.

\subsection{Model Reliability and Interpretability}
In clinical contexts, reliability and interpretability are prerequisites for adoption. 
However, VLMs frequently operate as black-box systems with limited transparency in their decision-making pathways~\cite{muhammad2024unveiling}. 
This challenge is exacerbated in adaptation strategies involving VLM-based supervision, such as generated images or text for data augmentation, where content validity is difficult to ensure. 
Without rigorous medical constraints, generative models may introduce implausible or misleading features, which can propagate erroneous inductive biases during training or inference~\cite{clusmann2025prompt}.

The lack of explainability further hampers trust. 
Clinicians require not only accurate predictions but also clinically meaningful justifications aligned with observable imaging evidence and domain knowledge. 
Yet, current VLM outputs rarely provide such traceability, especially in tasks like report generation and VQA~\cite{clusmann2025prompt}. 

This issue particularly limits the clinical viability of Supervision Enhancement and Direct Deployment strategies, where generated outputs are used as training signals or directly shown to end-users. The lack of semantic control and clinical explanation can introduce bias and hinder safe deployment in real-world workflows.


\subsection{Modality Misalignment}
The effectiveness of VLMs in medical imaging heavily depends on the precise alignment between visual and textual modalities. 
However, most existing feature extraction and fusion mechanisms, such as dual-encoder architectures or naive matrix operations, fail to capture the complex semantic correlations between anatomical structures and domain-specific language. 
While attention-based methods, particularly cross-attention, offer more expressive inter-modal modeling, they suffer from computational inefficiencies and scalability issues when applied to high-resolution medical images~\cite{li2023snapfusion}.

Moreover, fusion misalignment is particularly problematic in downstream tasks requiring fine-grained interaction between modalities. 
For instance, inaccurate semantic grounding between a textual cue (e.g., ``peripheral spiculated nodule") and its corresponding image region can degrade performance in localization-sensitive tasks like segmentation, object detection, or visual grounding~\cite{liu2024g2d}.

These issues directly challenge the robustness of the Feature Extraction strategy, where successful cross-modal fusion is critical. Misaligned representations can result in unstable multi-modal embeddings and severely impair fine-grained reasoning required for medical applications.

\subsection{Cross-Modal Adaptation}
General-domain VLMs, pretrained on large-scale natural image-text corpora, often fail to generalize to the medical domain due to substantial domain shifts~\cite{azad2023foundational}. 
Medical imaging spans a wide range of modalities with unique spatial characteristics and diagnostic priors, which differ drastically from natural scenes~\cite{lee2024read}. 
Simultaneously, medical text is highly structured, information-dense, and context-dependent, which are characteristics not typically represented in general language corpora~\cite{monajatipoor2024medical}.

Existing adaptation techniques, including task-specific fine-tuning or prompt-based transfer learning, only partially mitigate this domain gap. 
Fine-tuning often requires annotated medical data that are expensive to obtain, while prompt tuning lacks sufficient flexibility to encode detailed domain knowledge or spatial cues~\cite{hussein2024promptsmooth,windsor2023vision}. 
These limitations are particularly pronounced in rare disease settings or low-resource scenarios where generalization across underrepresented cases is critical.

This domain gap severely constrains the effectiveness of the Task-specific Fine-tuning strategy, which require either precise tuning with labeled data or strong prior alignment. In data-scarce or specialty domains, both adaptation mechanisms struggle to maintain performance.

\subsection{Clinical Deployment Constraints}
Despite growing interest in deploying VLMs for real-world clinical tasks, current systems face significant deployment barriers. 
Most state-of-the-art VLMs are computationally intensive, requiring substantial GPU memory and processing power, which restricts their applicability in resource-limited clinical environments~\cite{liu2023m,ates2025dcformer}. 
While techniques such as model compression~\cite{dantas2024comprehensive}, pruning~\cite{he2024rethinking}, can reduce runtime complexity, they often trade off predictive performance, especially on specialized medical subtasks.

Furthermore, many clinical applications, such as emergency diagnostics, telemedicine triage, or real-time decision support, demand low-latency, interactive AI systems. 
However, the inference time of large VLMs, especially in multi-step pipelines involving report generation or multi-modal reasoning, frequently exceeds acceptable clinical thresholds. 
These deployment challenges are especially critical for the Direct Deployment Strategy. Achieving high-throughput, low-latency VLMs that retain diagnostic fidelity and interpretability is essential for enabling their practical integration into healthcare workflows. 

\section{Future Directions}
\label{sec:future directions}

In this section, we explore future research directions for applying vision-language models (VLMs) to medical image analysis, grounded in the aforementioned challenges. The discussion encompasses several key aspects, including data synthesis, cross-domain adaptation in low-shot scenarios, interpretability improvement, and practical deployment in real-world clinical scenarios.

\subsection{Automatic Labeling and Data Synthesis}
To address the scarcity of labeled data in medical imaging, future research should explore leveraging medical VLMs (\eg, LLaVA-Med~\cite{li2024llava}, Qilin-Med-VL~\cite{liu2023qilin}) to generate pseudo-labels and synthetic clinical narratives. These auto-generated annotations can serve as weak supervision for training medical image analysis models, especially in domains with limited annotations. Moreover, feed synthetic reports into generative image models facilitates the creation of diverse, multi-modal datasets~\cite{bahani2024enhancing}. This is particularly useful for representing rare diseases and underrepresented modalities. Future efforts may also involve refining synthetic data through adversarial training or domain adaptation~\cite{lee2023vision}, improving realism and bridging gaps between synthetic and real distributions.

\subsection{Developing Secure Data Sharing Protocols}
To address the fragmentation of medical data across institutions, a promising research direction lies in the design of collaborative learning frameworks that support secure and efficient data sharing~\cite{imani2019framework}. Federated learning~\cite{zhu2024stealing} enables decentralized model training without exchanging raw data, allowing institutions to contribute while retaining control over their data. Beyond security, effective framework design must consider system scalability, communication efficiency, and interoperability across heterogeneous infrastructures. Techniques such as secure aggregation and differential privacy safeguard sensitive information, while encrypted computation reinforces trust, paving the way for scalable, cross-institutional collaboration.

\subsection{Fine-grained Cross-modal Alignment}
Accurate alignment of visual and textual features is vital for high-performing medical VLMs. Clinical texts vary in density and structure, requiring adaptable strategies for multi-level alignment. Techniques like fine-grained region-to-text attention mechanisms and semantic embedding refinement can improve interpretability and diagnostic relevance. Subsequent research could integrate anatomical priors or knowledge-based constraints to guide latent space alignment~\cite{jiang2023anatomical}, allowing models to better recognize subtle pathologies. Additionally, dynamic attention mechanisms~\cite{dai2021dynamic} that adjust based on text complexity can help improve robustness. These approaches can collectively enhance semantic coherence between images and reports, boosting clinical applicability. Exploring new alignment losses or contrastive learning may further benefit performance.

\subsection{Integrating Knowledge Graphs with VLMs}
Integrating structured knowledge like SNOMED CT~\cite{donnelly2006snomed} or UMLS~\cite{bodenreider2004unified} into VLMs can enhance explainability and consistency. Knowledge graphs anchor features to known clinical concepts, supporting interpretable supervision. Further investigation may focus on graph-guided learning using attention mechanisms~\cite{ma2024focus} or loss functions based on graph topology~\cite{chen2024knowledge}. These techniques can help reduce hallucinations and improve model robustness. Structured relationships also assist in disambiguating similar terms and detecting rare conditions, promoting richer and more reliable diagnostics.

\subsection{Low-resource Cross-modal Adaptation}
In data-limited scenarios, it is challenging for VLMs to achieve efficient adaptation. Semi-supervised learning~\cite{li2024vclipseg, li2024textmatch}, meta-learning~\cite{leng2024self}, and few-shot techniques~\cite{liu2024few, zhu2024toward} offer promising solutions for generalization across tasks with minimal supervision. Methods such as Contrastive learning frameworks, such as SimCLR~\cite{chen2020simple} or MoCo~\cite{he2020momentum}, can be adapted to medical VLMs by maximizing agreement between cross-modal embeddings of the same case while pushing apart unrelated instances. Furthermore, pretrained VLMs can be fine-tuned using prototypical networks or gradient-based meta-learners~\cite{lee2018gradient} to achieve rapid adaptation in few-shot settings. These methods enhance model generalization in resource-constrained environments and support deployment in rare disease applications.

\subsection{Lightweight and Modular Medical VLMs}
Scalability in medical VLMs demands lightweight, modular architectures. A promising direction is to explore parameter-efficient tuning methods like LoRA~\cite{hu2021lora}, adapters~\cite{houlsby2019parameter}, and prompt tuning~\cite{lester2021power} to minimize retraining needs while preserving performance. These modular components can be flexibly deployed for downstream medical image analysis tasks across resource-constrained settings. Edge-friendly designs, such as quantization and task-specific pruning, further reduce model size without sacrificing accuracy. Combining these ideas with continuous learning frameworks~\cite{sun2024continually} would allow adaptive model updates without retraining from scratch. Such modular systems are key for practical, sustainable VLM deployment in diverse clinical environments. 

\subsection{Knowledge Distillation}
To complement modular optimization strategies, knowledge distillation focuses on transferring representational capacity from large VLMs to compact student models. Beyond parameter reduction, distillation enables semantic preservation by aligning intermediate features~\cite{wang2024feature}, attention maps~\cite{elnoor2025vi}, or cross-modal~\cite{wang2022multimodal} relations between teacher and student networks. In medical contexts, structure-aware distillation techniques~\cite{yang2024rebalanced} can emphasize clinically relevant features such as lesion boundaries or anatomical landmarks. Relation-based~\cite{park2019relational} and contrastive~\cite{tian2019contrastive} distillation approaches further ensure semantic consistency in multi-modal embeddings. Unlike pruning or quantization, distillation retains interpretability~\cite{han2023impact} while significantly reducing inference latency. These methods are particularly effective in maintaining diagnostic performance when deploying models to point-of-care or mobile imaging systems.

\section{Conclusion}
\label{sec:conclusion}

In this survey, we provide a systematic review of task-specific adaptation strategies for vision-language models (VLMs) in medical image analysis. We first outline the training paradigms of medical VLMs. Then, five representative adaptation strategies are categorized and analyzed across key clinical tasks such as segmentation, diagnosis, and report generation. While VLMs demonstrate strong potential in enhancing multi-modal understanding in medical image analysis, challenges persist in domain alignment, limited annotated data, and insufficient model interpretability. To support clinical deployment, future efforts should focus on scalable data construction, aligned cross-modal representation learning, and developing lightweight deployment strategies. This review aims to offer a structured perspective on medical VLM adaptation and provide guidance for future research and real-world applications of VLMs in medical image analysis.



\bibliographystyle{elsarticle-num} 
\bibliography{ref}





\end{document}